\documentclass[reprint,aps,prb,twocolumn,groupedaddress,amsmath,amssymb,longbibliography,floatfix]{revtex4-2}
\usepackage{graphicx}
\usepackage[colorlinks=true,citecolor=blue,
            linkcolor=blue,urlcolor=blue,bookmarks=true]{hyperref}
\usepackage[utf8]{inputenc}
\usepackage[english]{babel}
\usepackage[final]{microtype}
\microtypesetup{protrusion=true,expansion=true}
\usepackage{amsmath}
\usepackage{xcolor}
\usepackage{soul}
\usepackage{mathbbol}
\usepackage{multirow}
\usepackage{physics}
\usepackage{listings}
\definecolor{codekw}{RGB}{0,0,180}
\definecolor{codestr}{RGB}{163,21,21}
\definecolor{codecomment}{RGB}{0,128,0}
\definecolor{codebg}{RGB}{248,248,248}
\lstdefinestyle{pythonstyle}{
  language=Python,
  basicstyle=\ttfamily\footnotesize,
  keywordstyle=\color{codekw}\bfseries,
  stringstyle=\color{codestr},
  commentstyle=\color{codecomment}\itshape,
  backgroundcolor=\color{codebg},
  showstringspaces=false,
  breaklines=true,
  columns=fullflexible,
  keepspaces=true,
  upquote=true,
  xleftmargin=4pt,
  framexleftmargin=2pt,
}
\makeatletter
\renewcommand{\fnum@figure}[1]{\textbf{FIG.~\thefigure.~}}
\makeatother

\begin{document}

\title{VeloxQ: \texorpdfstring{\\}{ }A Fast and Efficient QUBO Solver}

\author{J. Paw\l{o}wski}
\affiliation{Institute of Theoretical Physics, Faculty of Fundamental Problems of Technology, Wroc\l{a}w University of Science and Technology, 50-370 Wroc\l{a}w, Poland}
\affiliation{Quantumz.io Sp.\;z\;o.o., Puławska 12/3, 02-566 Warsaw}

\author{J. Tuziemski}

\affiliation{Quantumz.io Sp.\;z\;o.o., Puławska 12/3, 02-566 Warsaw}

\author{P. Tarasiuk}
\affiliation{Quantumz.io Sp.\;z\;o.o., Puławska 12/3, 02-566 Warsaw}

\author{H. Louzada}
\affiliation{Quantumz.io Sp.\;z\;o.o., Puławska 12/3, 02-566 Warsaw}

\author{R. Adamski}
\affiliation{Quantumz.io Sp.\;z\;o.o., Puławska 12/3, 02-566 Warsaw}

\author{K. Hendzel}
\affiliation{Quantumz.io Sp.\;z\;o.o., Puławska 12/3, 02-566 Warsaw}

\author{Ł. Pawela}
\affiliation{Institute of Theoretical and Applied Informatics, Polish Academy of Sciences, Ba{\l}tycka 5, 44-100 Gliwice, Poland}
\affiliation{Quantumz.io Sp.\;z\;o.o., Puławska 12/3, 02-566 Warsaw}

\author{B. Gardas}
\affiliation{Institute of Theoretical and Applied Informatics, Polish Academy of Sciences, Ba{\l}tycka 5, 44-100 Gliwice, Poland}
\affiliation{Quantumz.io Sp.\;z\;o.o., Puławska 12/3, 02-566 Warsaw}

\begin{abstract}
    We introduce VeloxQ, a fast solver for Quadratic Unconstrained Binary Optimization (QUBO)
    problems, which are central to many real-world optimization tasks.
    Unlike approaches that depend on emerging quantum hardware, VeloxQ can be deployed on conventional computing infrastructure.
    We benchmark VeloxQ against state-of-the-art QUBO solvers from several families.
    These include quantum annealers, specifically D-Wave's Advantage and Advantage2 platforms;
    the digital-quantum BF-DCQO algorithm for Higher-Order Unconstrained Binary Optimization (HUBO)
    developed by Kipu Quantum; physics-inspired algorithms including Simulated Bifurcation,
    Parallel Annealing, and tropical tensor networks; and conventional methods including CPLEX,
    brute force, BEIT's Chimera solver, and Branch-and-Bound variants.
    The benchmark suite covers native quantum-annealer topologies, embedded all-to-all instances,
    HUBO-derived instances, planted-solution instances, certified-solver regimes, and dense
    Branch-and-Bound test cases.
    Across the benchmark suite, VeloxQ delivers competitive solution quality and runtime, and in several regimes outperforms the compared solvers.
    VeloxQ also demonstrates strong scalability. Among the solvers considered in this study, it was the only method we could run on the largest sparse instances within 
    our computational budget, including problems with up to $10^{8}$ sparsely connected variables.
    These findings position VeloxQ as a competitive and practical tool for tackling large-scale QUBO/HUBO problems, offering a practical alternative to existing
    quantum and classical optimization methods.

\end{abstract}

\maketitle

\section{Introduction}

The increasing complexity of real-world optimization problems poses a serious challenge to state-of-the-art methods.
The growing size and complexity of industry-relevant instances may soon limit the performance of well-established exact and heuristic approaches.
Recent progress in emerging computational paradigms, including quantum computing and physics-inspired algorithms, has opened new opportunities for the development of optimization methods that may outperform conventional techniques on selected classes of QUBO problems.
For example, simulating the dynamics of certain classical systems has recently proven to be a promising heuristic approach to QUBO optimization~\cite{Goto2016,Goto2019,Jiang2023}.
In contrast to other physics-inspired optimization algorithms such as Simulated Annealing~\cite{SA}, these approaches can take advantage of conventional parallel hardware.

Here we present VeloxQ~\cite{veloxq}, a QUBO solver based on a novel physics-inspired methodology.
Unlike approaches that rely on further advances in quantum hardware, VeloxQ can fully exploit conventional computing hardware today.
In this sense, VeloxQ bridges established and emerging approaches to optimization and may also serve as a practical component of hybrid quantum-classical workflows, although such integrations are outside the scope of the present paper.

This document is intended as a technical white paper aimed at practitioners considering QUBO solvers for production use, rather than as a peer-reviewed journal submission.
Its aim is to provide an extensive set of benchmarks that characterizes key features of VeloxQ, including solution quality and runtime.
To place VeloxQ in the broader QUBO landscape, we compare it with a wide range of competing approaches, including the commercial optimization suite CPLEX, quantum-analog solvers offered by D-Wave, the digital-quantum optimization algorithm developed by Kipu Quantum~\cite{romero2024}, and an optimized implementation of the brute-force QUBO solver~\cite{Jalowiecki_2021,Jalowiecki_2023,BF}.
The comparison also covers solvers with ground-state certification, including the BEIT QUBO solver~\cite{BEITQsolver,BEIT_AWS} and a physics-inspired algorithm based on tropical tensor networks~\cite{Liu2021}, as well as heuristic and classical methods such as parallel annealing~\cite{Jiang2023}, simulated bifurcation~\cite{Goto2016}, and modern Branch-and-Bound variants~\cite{Chalkis2023,Quantagonia2023}.

\begin{figure}[!tbp]
    \centering
    \includegraphics[width=\linewidth]{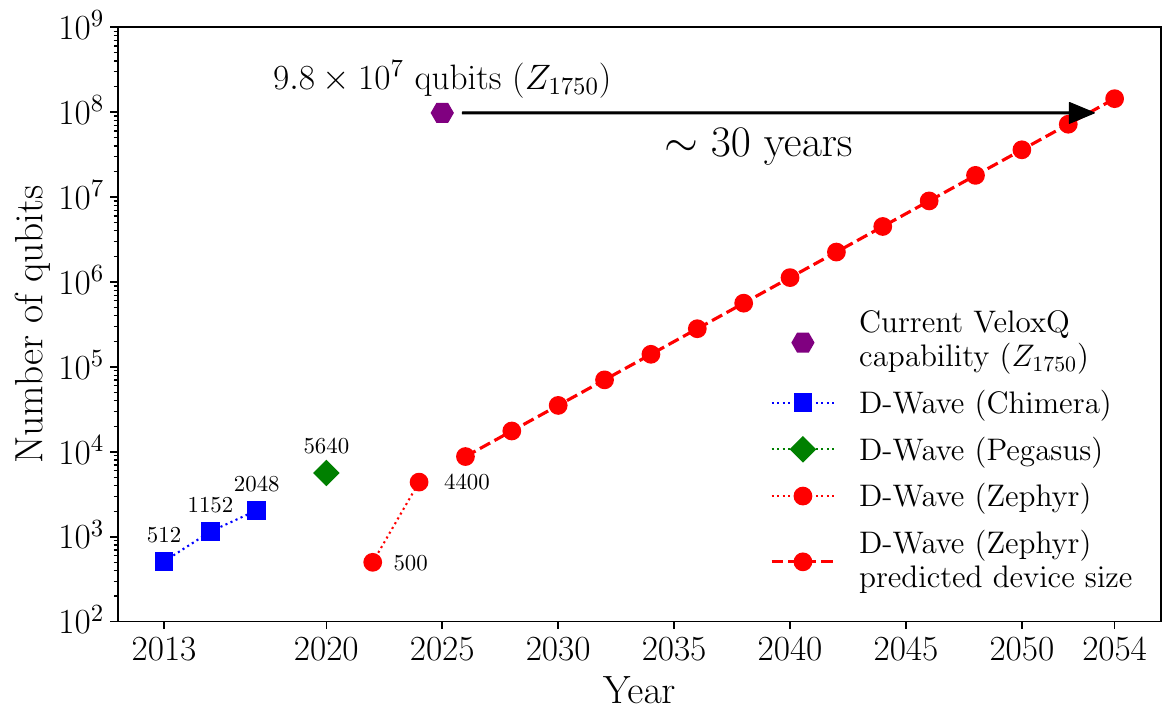}
    \caption{Comparison of past and current D-Wave quantum-annealer qubit counts with an illustrative extrapolation to the \(Z_{1750}\) instance solved by VeloxQ, containing \(\sim 9.8\times 10^7\)
        variables (purple hexagon). Under a simple doubling-every-two-years extrapolation estimated from historical release data, a device capable of
        natively handling such instances would be expected around 2054.
    }
    \label{fig:dwave_roadmap}
\end{figure}

The benchmark set includes instances that are favorable to the solvers compared with VeloxQ.
This choice allows us to identify their strengths and reduces the risk that VeloxQ performance appears favorable only because of biased instance selection.
The three central claims tested below are solution quality, runtime, and scalability.
Across these benchmarks, VeloxQ is competitive with the compared solvers in solution quality and runtime, while its main distinction is scalability: among the methods included in this study, it is the only one that we were able to run successfully on the largest sparse instances considered here.
As in several of the compared methods, however, VeloxQ results are not certified, \emph{i.e.}, the returned solution is not guaranteed to be optimal.
For example, VeloxQ can solve an optimization problem on the Zephyr graph $Z_{1750}$, which involves more than $9.8 \times 10^{7}$ variables.
As an illustrative scale comparison, Fig.~\ref{fig:dwave_roadmap} shows a simple extrapolation from past D-Wave device sizes to a native graph of that magnitude; this estimate should be interpreted only as a rough indication of scale.
A similar scale separation appears in the comparison with the digital quantum algorithm developed by Kipu Quantum, for which VeloxQ can handle an instance with $10^{8}$ variables.

There are notable emerging QUBO solution technologies that we do not include in this comparison.
Memcomputing is a promising approach to QUBO optimization~\cite{Sheldon_2019}; however, we were not able to access a solver based on this methodology.
Considerable effort has also been devoted to coherent Ising machines, both from the hardware perspective~\cite{Si2024} and in the theoretical study of the relation between binary solutions and their continuous relaxations~\cite{Yamamoto_2020,Wang2023}.
We decided not to include coherent Ising machines because their behavior is similar to that of the simulated bifurcation algorithm.
We also implemented an annealer based on Hamiltonian Monte Carlo~\cite{Wang2025-zk}, but the initial results were not satisfactory and we therefore did not pursue an extensive benchmark.
Finally, approaches based on classical thermodynamics, such as quadratic programming enhanced by thermodynamic linear algebra subroutines~\cite{bartosik2024, Aifer2024}, were not considered because they require specialized hardware that is currently unavailable and do not solve binary problems natively.

To make the paper self-contained, Sec.~\ref{sec:qubo} introduces the definition of QUBO problems, their relation to the Ising model, and higher-order unconstrained binary optimization.
Sec.~\ref{sec:qubo_solvers} reviews state-of-the-art approaches to solving QUBO problems.
The benchmark sections are organized as follows.
Sec.~\ref{sec:benchmarks_dwave} presents benchmarks against D-Wave quantum annealers.
Sec.~\ref{sec:benchmarks_hubo} compares VeloxQ with the digital quantum algorithm developed by Kipu Quantum.
Sec.~\ref{sec:benchmarks_certification} discusses benchmarks against solvers with certified solutions, including the BEIT solver~\cite{BEITQsolver,BEIT_AWS} and a tropical tensor network approach.
Sec.~\ref{sec:physicsInpiredAlgorithms} presents results against physics-inspired algorithms, namely parallel annealing and simulated bifurcation.
Sec.~\ref{sec:benchmarks_BB} compares VeloxQ with a refined Branch-and-Bound (B\&B) algorithm proposed by Quantagonia~\cite{Quantagonia2023}.
We summarize our findings in Sec.~\ref{sec:summary}.
Each benchmark addresses a different problem: instances native to quantum hardware, impact of embedding overhead, reduction of higher-order problems to QUBO,
cost of certified solution and performance on planted-solution instances.

All benchmark instances, generation scripts, plotting scripts, and relevant data
are provided in the separate online repository~\cite{repository}. While VeloxQ is a proprietary solver, access is
provided via the open Python~API \texttt{veloxq\_sdk}~\cite{sdk}. Sec.~\ref{sec:dwave_sdk_comparison}
discusses the API in more detail, including a comparison against the D-Wave~API.

\section{Quadratic Unconstrained Binary Optimization}
\label{sec:qubo}
QUBO problems constitute an important class of optimization problems with applications in many areas ranging from operations research to portfolio optimization~\cite{Kochenberger2014}. They encode the task of finding a binary vector $\vb{x} \in \{0,1 \}^N$
that minimizes a quadratic function $Q(\vb{x})$,
\begin{equation}
    Q(\vb{x}) = \sum_{i\leq j} Q_{ij} x_i x_j .
    \label{eq:qubo_model}
\end{equation}
The optimization problem encoded in QUBO can be equivalently formulated in terms of the Ising model. The latter originates from statistical physics, and describes the energy associated with a particular configuration of discrete variables
(spins) $\vb{s} \in \{-1, 1\}^N$:

\begin{equation}
    H(\vb{s}) = \sum_{i<j}  J_{ij} s_i s_j + \sum_{i} h_i s_i.
    \label{eq:ising_model}
\end{equation}
The transformations between QUBO and Ising formulations are presented in Appendix~\ref{app:QUBOIsingconversion}. The Ising-QUBO relation opens the possibility of tackling optimization problems with analog quantum devices such as quantum annealers,
as well as gate-based quantum computers~\cite{Abbas2024}. Despite the equivalence, there are cases (e.g., random instances) in which applying a QUBO solver to a QUBO formulation of the Ising problem may perform poorly~\cite{Boettcher}.

In general, QUBO belongs to the class of NP-hard problems, and many NP problems can be embedded into QUBO~\cite{Glover2022}, such as
graph coloring~\cite{lewis2021guide}, max-cut~\cite{Commander2009}, the traveling salesman problem~\cite{Hoffman2013}, or Boolean satisfiability~\cite{Eggersg2012}, as well as
Higher-Order Unconstrained Binary Optimization problems~\cite{dattani2019}, whose cost function includes multi-variable terms of order $\mathcal{P}$:
\begin{equation}
    P(\vb{x}) = \ \sum_{\substack{ i_1,\cdots,i_N \\ i_1 + \cdots + i_N \le \mathcal{P}}} \ a_{i_1 \dots i_N} x_1^{i_1} \dots x_N^{i_N} \, .
\end{equation}
Third-order HUBO problems will be used to benchmark VeloxQ against proposed quantum algorithms capable of solving such HUBO problems natively.

\section{State-of-the-Art QUBO Solvers}
\label{sec:qubo_solvers}

For the purpose of this paper, we divide solvers capable of handling QUBO problems into the following categories:
classical, analog-quantum, and digital-quantum solutions, as well as physics-inspired algorithms.

The first group consists of solvers that run on conventional hardware. CPLEX~\cite{CPLEX} and Gurobi~\cite{gurobi} have been widely used in industry for optimization
problems, including QUBO problems. They implement both exact and heuristic solving methods~\cite{BIER18}. Linear integer programming solvers can also
be used to handle QUBO problems, due to the existence of reformulations of QUBO as a linearly constrained binary problem. The performance of general-purpose classical solvers such as those mentioned above
depends strongly on the problem structure and its size. Some classical solvers are designed to handle problems with a specific structure, e.g., the BEIT QUBO solver~\cite{BEITQsolver,BEIT_AWS}, which accepts instances of Chimera graph
topology and restricted size. There are also attempts to utilize deep reinforcement learning to solve QUBO problems~\cite{Fan2023,Boettcher2023}.

Due to the connection between QUBO problems and Ising models, quantum annealers have been employed to solve QUBO problems~\cite{Crosson2021}.
The main limitation of this approach is the number of qubits available on a quantum annealer,
and its topology, which restricts the class of problems that can be natively implemented on a device. Problem instances not matching the topology of the device require embedding that
introduces computational overhead. Whether quantum annealers offer a computational advantage over classical methods is still debated~\cite{Villanueva_2023, Lidar2025, Pawlowski2025}.
Specific use cases continue to be investigated, including protein folding~\cite{Folding} and grid cost allocation in electricity markets~\cite{gridcostallocationpeertopeer}.
Quantum annealers also serve as valuable scientific tools for studying the quantum Ising model~\cite{verresen2023quantumisingmodel,king2024computationalsupremacyquantumsimulation}.

In the context of gate-based quantum computers there exists a family of optimization algorithms exploiting subroutines such as Quantum Phase Estimation or the Grover algorithm~\cite{Abbas2024},
and there are cases of quantum algorithms that exhibit super-polynomial speedup over classical algorithms for a certain class of optimization problems~\cite{EisertOptimization}. However, those algorithms remain theoretical proposals
since current quantum hardware, despite recent progress~\cite{Acharya2024}, has not yet reached the scale and reliability typically required for commercially relevant combinatorial optimization problems.
On the other hand, the so-called Variational Quantum Algorithms (e.g., Quantum Approximate Optimization Algorithm~\cite{QAOA}) are proposed as quantum heuristics that can be run on available devices. However, in most cases it remains unclear
whether such algorithms lead to an advantage over the best-known classical algorithms. For a recent overview of the field see~\cite{Abbas2024}.

There is also a group of physics-inspired algorithms that utilize conventional computing technology. Arguably the best-known algorithm
belonging to this group is Simulated Annealing, which was introduced in the 1980s~\cite{SA}, and is still being studied, e.g.,
to develop optimized versions for particular use cases~\cite{Isakov_2015}. Recently developed physics-inspired algorithms include Simulated Bifurcation~\cite{Goto2016},
tensor-network methods~\cite{dziubyna2024}, and classical algorithms inspired by quantum annealing~\cite{Jiang2023,Zeng2024}. It is also interesting to note that many algorithms in convex optimization, such as Alternating Direction
Method of Multipliers~\cite{ADMM}, can be formulated in terms of dynamical systems~\cite{franca2018admmacceleratedadmmcontinuous,Fran_a_2021,Fran_a_2023}.

\section{Benchmark methodology}
\label{sec:benchmark_methodology}

This section collects the conventions used across the benchmarks. The aim is to make the comparisons readable and reproducible even though not all solvers expose the same timing information, hardware configuration, or optimality guarantees.

\subsection{Quality metrics and references}

When a certified optimum or planted ground state is available, we report the \emph{optimality gap}, defined relative to the known ground-state energy.
When no certified optimum is available, we report a \emph{reference gap} relative to the best reference solution obtained by a long-running auxiliary solver, usually SA or CPLEX.
For the candidate and reference configurations, the relative-energy gap is
\begin{equation}
    g(\vb{s}, \vb{s}_{\text{ref} }) = \frac{H(\vb{s}) - H(\vb{s}_{\text{ref}})}{|H(\vb{s}_{\text{ref}})|}.
\end{equation}
Both quantities use the same relative-energy formula, but only the optimality gap is a gap to a certified ground state.
Throughout the benchmark sections, ``reference gap'' is therefore used for comparisons against SA, CPLEX, or iterative VeloxQ references, whereas ``optimality gap''
is reserved for planted or otherwise certified ground-state references. We note that in principle, CPLEX is capable of producing certified solutions, but in practice,
a $\sim 10^3\,\mathrm{s}$ runtime limit was imposed on CPLEX runs, which in most cases led to non-certified solutions.

Reference solutions were obtained differently across regimes.
For D-Wave-native large instances, long-running GPU-based SA was used as the reference method.
For embedded dense instances and HUBO-derived instances where the size allowed it, CPLEX and/or SA provided reference energies.
For planted-solution families, the construction gives the ground-state energy directly.
For B\&B benchmarks on dense instances, the reference energy was approximated by invoking VeloxQ iteratively with an increasing number of internal states until the best energy repeated three consecutive times.

\subsection{Runtime accounting}

For VeloxQ, three runtime quantities are relevant.
The \emph{solver time} is backend GPU computation time.
The \emph{job time} includes solver computation and platform-side orchestration, but excludes client-side upload and download.
The \emph{wall time} is the full client-observed elapsed time, including upload, API~communication, platform overhead, computation, and result download.
Unless stated otherwise, VeloxQ runtimes in the benchmark figures report \emph{solver time}, including preprocessing and parameter autotuning.
The SDK~comparison in Sec.~\ref{sec:dwave_sdk_comparison} reports all three VeloxQ timing quantities explicitly.

For D-Wave~QPU runs, we use the \texttt{qpu\_access\_time} reported by the D-Wave~API rather than the annealing time of a single sample.
This includes QPU programming and repeated anneal-read cycles, including thermalization-related delays.
It does not include the full client-observed SDK~wall time, which is shown separately in Sec.~\ref{sec:dwave_sdk_comparison}.
For cloud-accessed services such as BEIT and D-Wave, queueing, request overhead, upload, and download may affect the wall time observed by a user and are not always separable from solver-side execution.

\subsection{VeloxQ settings and tuning}

Results denoted as `AutoTune VeloxQ' were obtained with the default configuration, without benchmark-specific knowledge supplied by the user.
AutoTune settings generally prioritize solution quality over the shortest possible runtime.
Results denoted as `Custom VeloxQ' use user-selected parameters to shift the trade-off toward shorter runtime, better solution quality, or a specific number of sampled low-energy states.
Where custom settings are used, the section text or figure caption states the relevant intent.
Other solvers were run using either the settings reported in the cited benchmark, default or recommended SDK settings, or the specific settings described in the corresponding benchmark section.
This creates a possible asymmetry in tuning effort, which is listed explicitly as a limitation in Sec.~\ref{sec:limitations_reproducibility}.

\subsection{Reproducibility of benchmarks}

Benchmark instances, generation scripts, plotting scripts, and relevant results data are provided in the accompanying online repository~\cite{repository}.
The generator scripts are intended to document the instance-generation procedure rather than serve as an exact regeneration pipeline for the full benchmark datasets,
which are available in the repository. Since the results are averaged over multiple realizations of random instances, they should also be statistically reproducible, even if the exact instances are not used.

VeloxQ itself is proprietary, but benchmark jobs can be reproduced through the open Python~API \texttt{veloxq\_sdk}~\cite{sdk}, subject to backend access and account permissions.
The SDK~comparison section~\ref{sec:dwave_sdk_comparison} used \texttt{veloxq\_sdk} version \texttt{1.0.0.dev16} and \texttt{dwave-ocean-sdk} version \texttt{9.3.0}.
While D-Wave was accessed through its API also in all other benchmarks, VeloxQ was launched directly on the backend rather than through the public API. This is not expected to affect reproducibility of the results,
since the API~exposes the relevant \emph{solver time} to the user.
Other external packages used include \texttt{CPLEX 22.1.1} with \texttt{CPLEX.jl 1.1.1}, Kerberos solver from \texttt{dwave-hybrid 0.6.15} and
Tropical Tensor Network implementation from the GitHub repository accompanying Ref.~\cite{Liu2021}. BEIT QUBO solver was accessed through its API in version \texttt{0.2.0}.
For the remaining physics-inspired solvers, namely parallel annealing, simulated bifurcation, and simulated annealing, as well as for Branch\&Bound,
custom GPU implementations were used.
All classical simulations were performed on a workstation equipped with dual Intel(R) Xeon(R) Platinum 8462Y+ CPUs, 4 NVIDIA H100 SXM GPUs, and 1 TB of RAM.
For relatively small benchmarks sets, with up to $\sim 5000$ variables, VeloxQ ran on a single GPU, while for larger problems, all four GPUs were used.

\section{Benchmarks against D-Wave quantum annealers}
\label{sec:benchmarks_dwave}

\begin{figure}[!tbp]
    \centering
    \includegraphics[width=\linewidth]{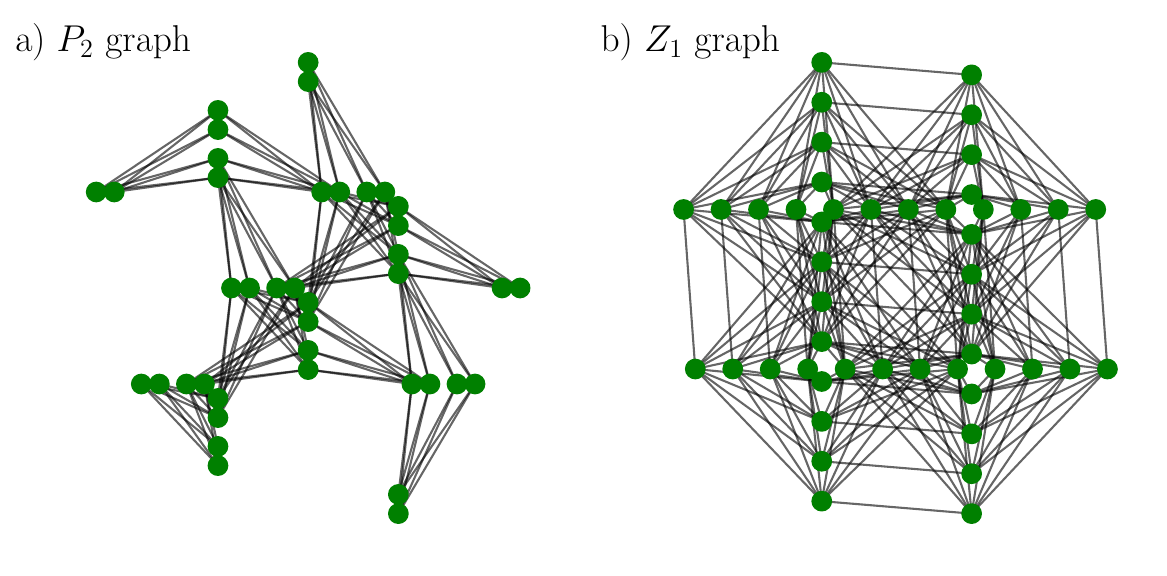}
    \caption{Illustration of the basic building blocks of the D-Wave's Pegasus~\cite{pegasus} and Zephyr~\cite{zephyr} topologies. \textbf{a)} Pegasus graph \(P_2\) on
        40 qubits, underlying current-generation D-Wave Advantage QPUs. This topology supports \(K_4\) and \(K_{6,6}\) graphs natively.
        \textbf{b)} Zephyr graph \(Z_1\) on $48$ qubits, forming the basis of the
        next-generation topology used in the Advantage2 and future devices. It enables direct embedding of \(K_4\) and \(K_{8,8}\) graphs.
    }
    \label{fig:dwave_topo}
\end{figure}

This section compares VeloxQ with state-of-the-art quantum annealing platforms, represented by D-Wave Advantage 6.4 and D-Wave Advantage2 1.12.
These quantum processing units (QPUs) are based on Pegasus \(P_{16}\) and Zephyr \(Z_{12}\) topologies, respectively (cf. Fig.~\ref{fig:dwave_topo}).
For most real-world problems, the target problem must be mapped onto the hardware graph through \textit{minor embedding}~\cite{membedding}, which can introduce significant overhead in both qubit usage and total solution time.
VeloxQ, by contrast, handles arbitrary problem topologies natively and does not require minor embedding.
This benchmark therefore tests two questions: how VeloxQ compares when the instance is native to the QPU topology, and how the comparison changes when embedding or hybrid decomposition becomes necessary.

Because of this structural difference, we begin with instances that are native to the D-Wave hardware and therefore do not require minor embedding.
Specifically, we solve QUBO problems with cost function given in Eq.~\eqref{eq:qubo_model}, where the interaction graph matches an undirected (sub)graph of the Pegasus or Zephyr topology.
The binary variables are placed on the vertices of the (sub)graph, and its edges correspond to couplings between variables.
For these benchmarks, we consider random QUBO instances with coupling matrix elements $Q_{ij}$ sampled uniformly from the interval \([-1,1]\).
To obtain reference solutions, we use Simulated Annealing~\cite{SA} and allow it to run much longer than the timescales on which both VeloxQ and D-Wave annealers operate, as illustrated, for example, in Fig.~\ref{fig:pegasus_native} d).

Real-world optimization problems, such as portfolio optimization~\cite{markowitz1967portfolio}, are often expressed by QUBO models whose number of variables considerably exceeds the number of qubits available on current quantum annealers.
To demonstrate the scalability of VeloxQ, we therefore also consider instances of the same structural type in a size range that is intractable for current D-Wave annealers.
For this regime, we use the \emph{dwave-hybrid} Python package and, in particular, the Kerberos hybrid solver, which uses a quantum annealer to process subproblems of the original instance~\cite{dwavehybrid}.
The benchmark set consists of Pegasus instances up to \(P_{150}\), with approximately $5\times10^{5}$ variables, and Zephyr instances up to \(Z_{150}\), with approximately $7.2\times10^{5}$ variables,
thus extending more than \(10^3\) times beyond the problem sizes accessible to current and near-future quantum annealers.
Reference solutions for these large-scale instances were again obtained using an efficient GPU-based implementation of the SA algorithm.

For each benchmark, we report AutoTune and Custom VeloxQ settings as defined in Sec.~\ref{sec:benchmark_methodology}.
\subsection{Pegasus topology}
\begin{figure}[!tbp]
    \centering
    \includegraphics[width=\linewidth]{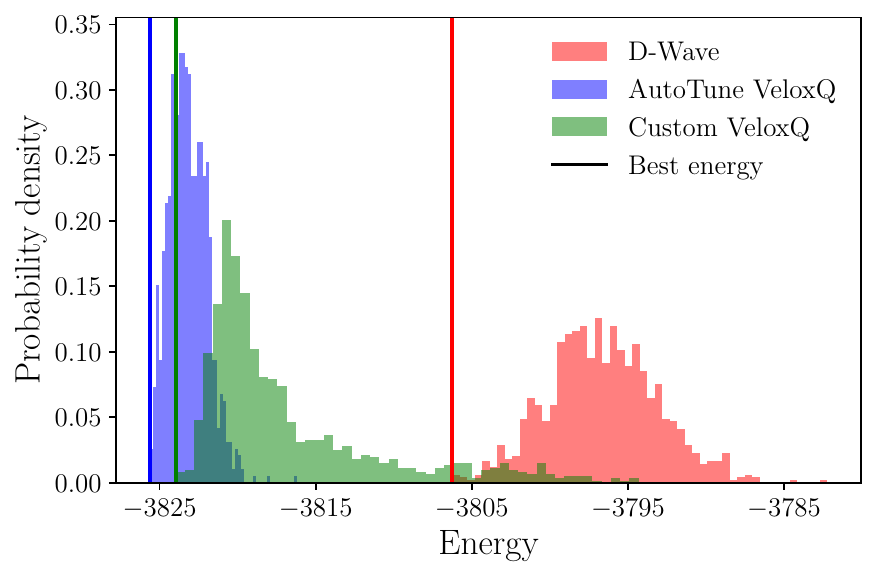}
    \caption{Low-energy spectra for a random QUBO problem on a \(P_{16}\) graph, with \(2^{10}\) states obtained from D-Wave Advantage (red), AutoTune VeloxQ (blue), and Custom VeloxQ (green). Custom VeloxQ is tuned for shorter runtime.}
    \label{fig:pegasus_native_spectra}
\end{figure}
To test statistical performance on a comprehensive set of problem sizes, we initially generated $20$ random
instances for each Pegasus subgraph of \(P_{16}\), namely \(P_2, P_3, \ldots, P_{16}\).
Since real QPUs are not perfect Pegasus graphs, these subgraphs were adapted to the actual hardware topology by removing missing nodes and couplers.
We report the reference gap and benchmark runtime using the conventions defined in Sec.~\ref{sec:benchmark_methodology}.
For these instances, the reference gap is computed relative to the long-running SA reference solution because certified ground states are unavailable.

Due to the inherently probabilistic nature of quantum measurements, D-Wave quantum annealers produce
an ensemble of samples, from which the best is usually selected as the final solution.
However, when the QPU is used as part of a hybrid solver that operates on selected subproblems, it can be beneficial to retain more than just the single best local result. This reduces the risk of getting trapped in a local minimum.
Although VeloxQ is a classical solver, it emulates this behavior and also produces low-energy spectra of solutions.
An example of such spectra for both VeloxQ and the D-Wave annealer is presented in Fig.~\ref{fig:pegasus_native_spectra}.
The VeloxQ distributions are shifted toward lower energies, and the AutoTune spectrum is more concentrated around the mean; the Custom spectrum, tuned for runtime, is broader but its lowest-energy state remains close to the best AutoTune configuration.
In the subsequent benchmarks, AutoTune VeloxQ and D-Wave annealers are set to produce \(2^{10}\) samples, and the best solution is used to calculate the reference gap.
Moreover, the D-Wave annealer is always configured to use an annealing time of $\tau = 100\,\mu\mathrm{s}$.
Both the reference gap and the runtime are averaged over problem realizations of a given graph size and over independent runs (see figure captions for details).

\begin{figure}[!tbp]
    \centering
    \includegraphics[width=\linewidth]{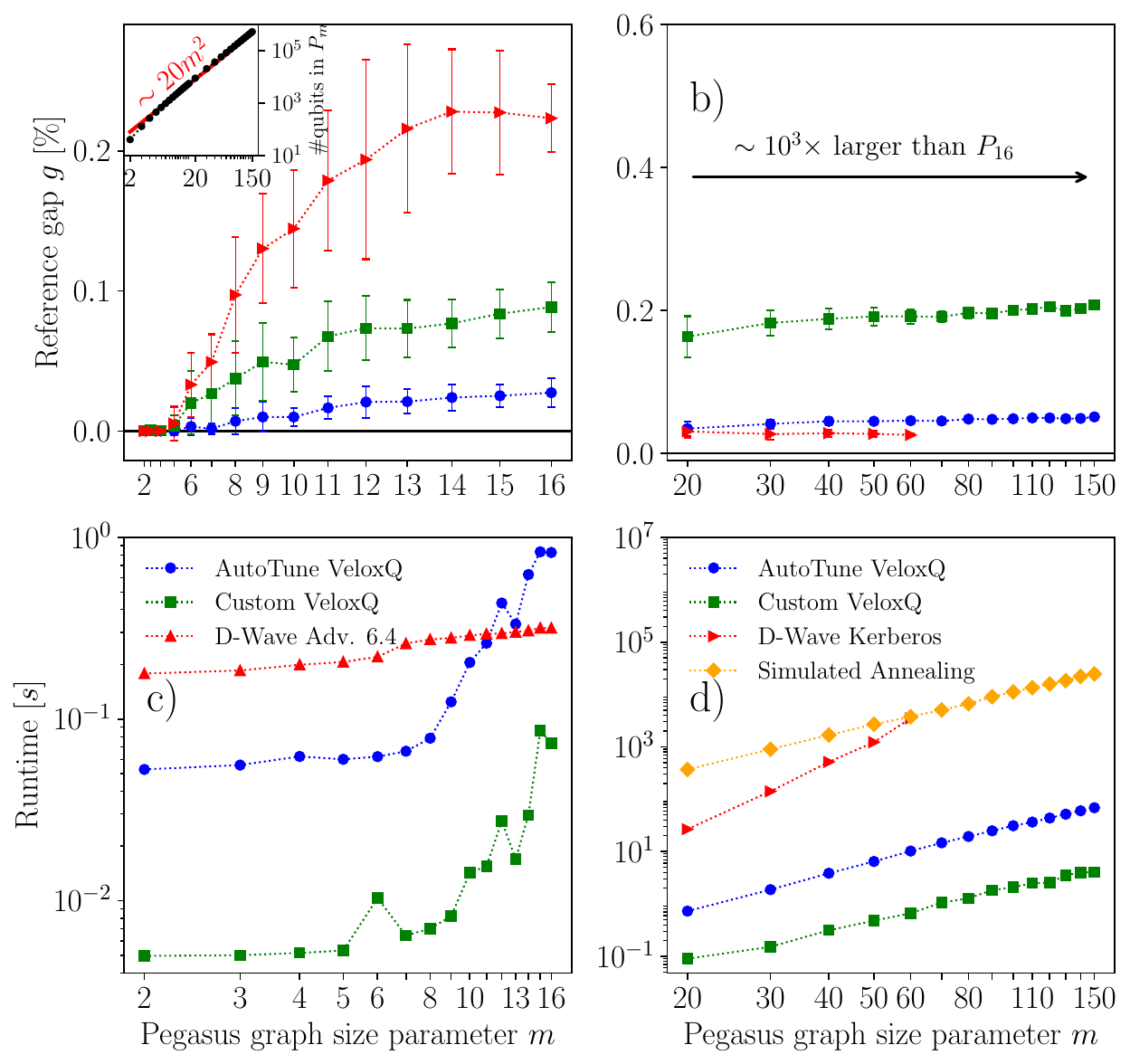}
    \caption{Performance of VeloxQ on problems with Pegasus topology, with D-Wave comparison where possible.
        Panels \textbf{a)}, \textbf{c)}: reference gap and reported runtime for \(P_{m\leq16}\) subgraphs of the D-Wave Advantage 6.4 Pegasus topology, averaged over $20$ random problems per graph size and $10$ independent runs. Error bars mark the standard deviation.
        The inset shows how the number of qubits scales with the size parameter of the graph.
        Panels \textbf{b)}, \textbf{d)}: the same quantities for larger Pegasus instances up to \(P_{150}\), averaged over $10$ random problems for $m\leq50$ and $5$ for $m>50$.
        The Kerberos solver was limited to instances up to \(P_{60}\), as for larger ones it ran out of memory on a machine with 1 TB of RAM.
    }
    \label{fig:pegasus_native}
\end{figure}

Results of these experiments are presented in Fig.~\ref{fig:pegasus_native}.
Panel \textbf{a)} shows that even with default settings (AutoTune VeloxQ), the solver obtains high-quality solutions close to the reference ones.
Although these instances are tailored to the D-Wave Advantage topology, we do not observe a solution-quality advantage for the annealer in this benchmark.

In terms of runtime, the quantum annealer is approximately \(5\times\) faster on the full QPU graph for instances close to the maximal available size.
By appropriately tuning VeloxQ parameters, we obtain similar or better runtime while maintaining better solution quality.

For runtime accounting on the D-Wave side, we do not use the annealing time alone as a proxy.
Instead, we account for all computation steps, including preparation and thermalization, as reflected in the \texttt{qpu\_access\_time} reported by the D-Wave~API.
The only practically relevant component that is omitted is the time related to the API~call itself, which we discuss separately in Sec.~\ref{sec:dwave_sdk_comparison}.

\begin{figure}[!tbp]
    \centering
    \includegraphics[width=\linewidth]{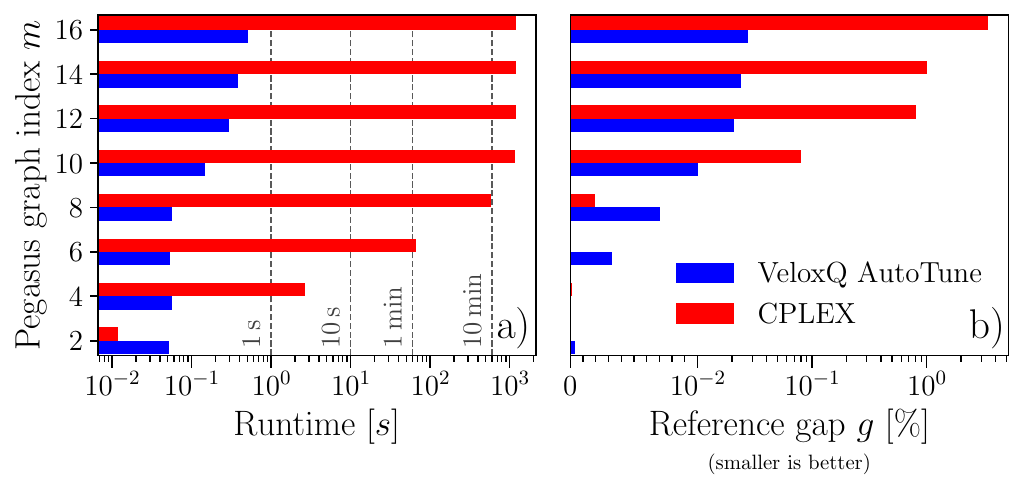}
    \caption{Comparison of VeloxQ with CPLEX on problems with Pegasus topology, the same
        as in Fig.~\ref{fig:pegasus_native}. Panel \textbf{a)} shows the average runtime, while
        panel \textbf{b)} shows the reference gap relative to the SA solution. CPLEX is faster than
        VeloxQ only for the smallest instance, which is expected due to the GPU overheads of VeloxQ.
        CPLEX runtime grows much faster with the problem size and reaches the time limit
        of 1000s already for \(P_{10}\), while VeloxQ runtime remains below $1s$.
    }
    \label{fig:pegasus_cplex}
\end{figure}

To investigate the regime beyond native QPU capabilities, we prepared 5 random Pegasus instances for size parameters up to \(m=150\).
The benchmark results are presented in Fig.~\ref{fig:pegasus_native} c) (reference gap) and d) (runtime).
VeloxQ consistently obtains high-quality solutions, matching the solution quality of the hybrid Kerberos solver while outperforming it in runtime by up to two orders of magnitude in this benchmark.
By adjusting VeloxQ parameters, the runtime can be further reduced by an order of magnitude at a relatively small cost in solution quality.
With the default parameters used here, the Kerberos solver was not able to produce solutions for instances larger than \(P_{60}\),
that is $\sim 85\times10^{3}$ variables, whereas VeloxQ produced solutions across the whole range of problem sizes considered, up to \(P_{150}\) with $\sim 500\times10^{3}$ variables.
A direct comparison of VeloxQ with the industry-standard CPLEX solver on the same Pegasus instances is shown in Fig.~\ref{fig:pegasus_cplex}: CPLEX is competitive only on the smallest cases and reaches its 1000\,s runtime limit already at \(P_{10}\), while VeloxQ stays well below 1\,s.

\begin{figure}[!tbp]
    \centering
    \includegraphics[width=\linewidth]{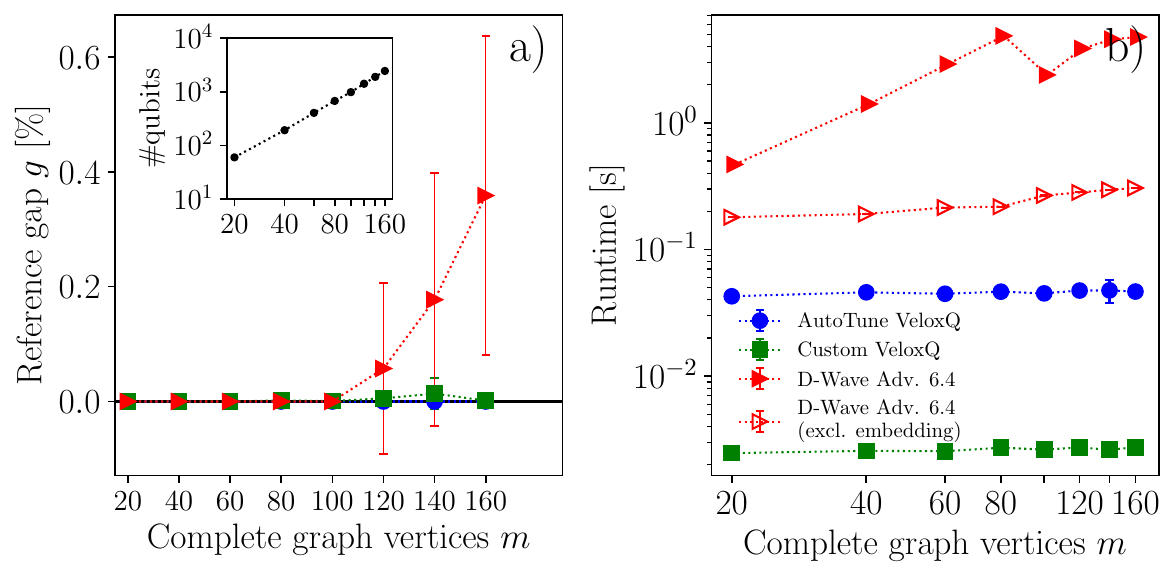}
    \caption{Performance of VeloxQ in comparison to the quantum annealer on problems with all-to-all connectivity.
        Panel \textbf{a)} shows reference gap relative to the CPLEX reference solution, averaged over 20 problem realizations and 10 runs. The inset shows the number of qubits needed to embed a complete graph into \(P_{16}\). Panel \textbf{b)} shows reported runtime; empty D-Wave symbols exclude embedding time and filled symbols include it.}
    \label{fig:pegasus_embedded}
\end{figure}

We also considered the opposite regime, namely problems with all-to-all connectivity, which are particularly challenging for quantum annealers because they require embedding.
We prepared random Ising models on complete graphs with up to
$160$ vertices, corresponding to $2536$ qubits after embedding.
The experiments were conducted, and the data were gathered in the same way as for instances with Pegasus topology.
The results are presented in Fig.~\ref{fig:pegasus_embedded}.
Even though there are specialized routines for clique embedding, the embedding overhead is associated with worsening reference gaps for the D-Wave Advantage sampler
beyond approx. $120$ variables. VeloxQ, on the other hand, solves these
problems with low reference gaps and short reported runtime, even in comparison to the annealer with the embedding time excluded.

\subsection{Zephyr topology}
\begin{figure}[!tbp]
    \centering
    \includegraphics[width=\linewidth]{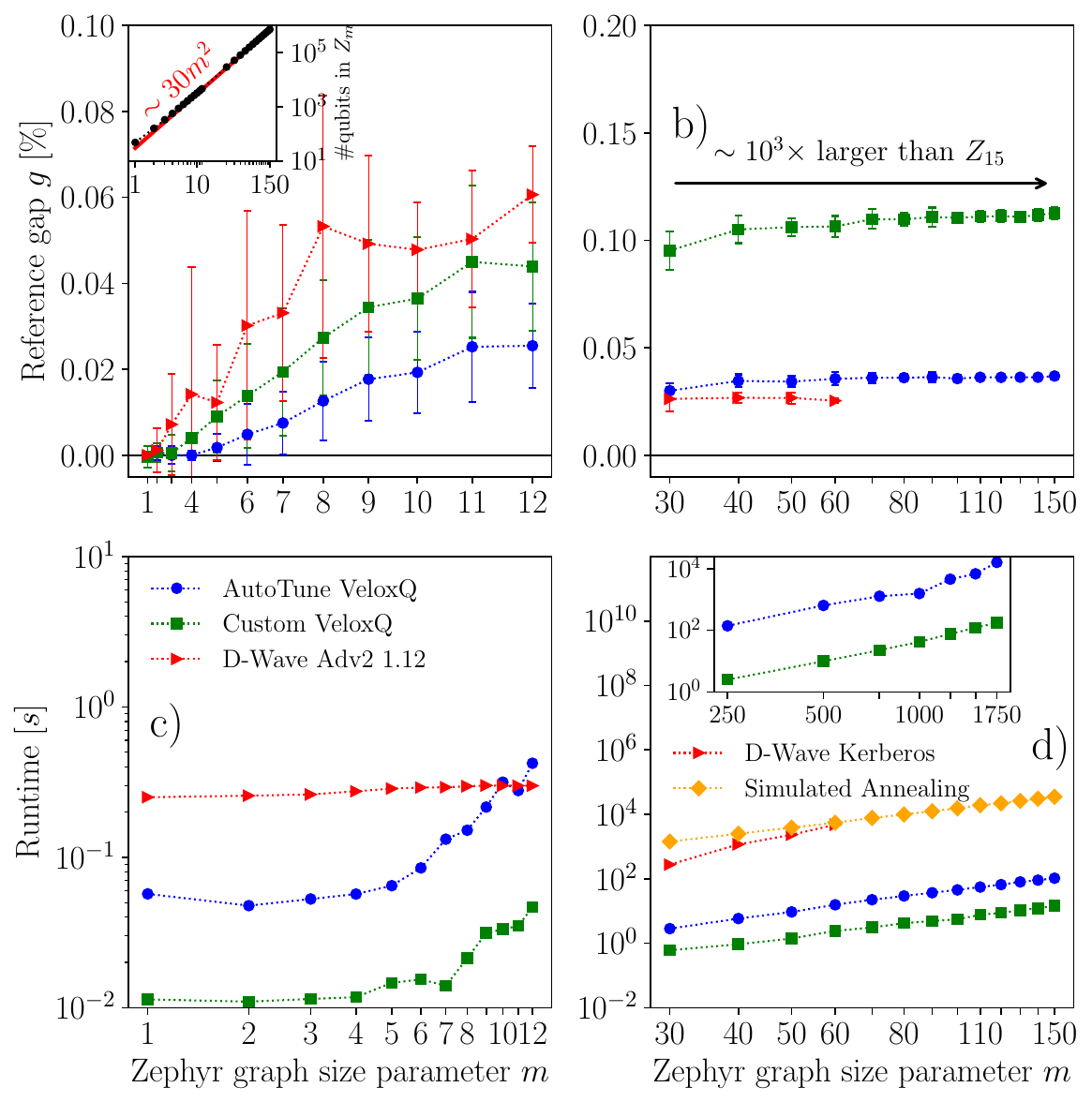}
    \caption{Performance of VeloxQ and D-Wave Advantage2 1.12 on Zephyr-topology problems.
        Panels \textbf{a)}, \textbf{c)}: reference gap and reported runtime for small native Zephyr instances. Both VeloxQ and D-Wave use $20$ random instances per graph size and $10$ independent runs per instance.
        The inset shows how the number of qubits scales with the size parameter of the graph.
        Panels \textbf{b)}, \textbf{d)}: the same quantities for larger Zephyr instances up to \(Z_{150}\), averaged over the available problem realizations for each graph size.
        The inset in panel \textbf{d)} shows VeloxQ runtime for larger Zephyr instances up to \(Z_{1750}\), for which no solution-quality claim is made.
    }
    \label{fig:zephyr_native}
\end{figure}

\begin{figure}[!tbp]
    \centering
    \includegraphics[width=\linewidth]{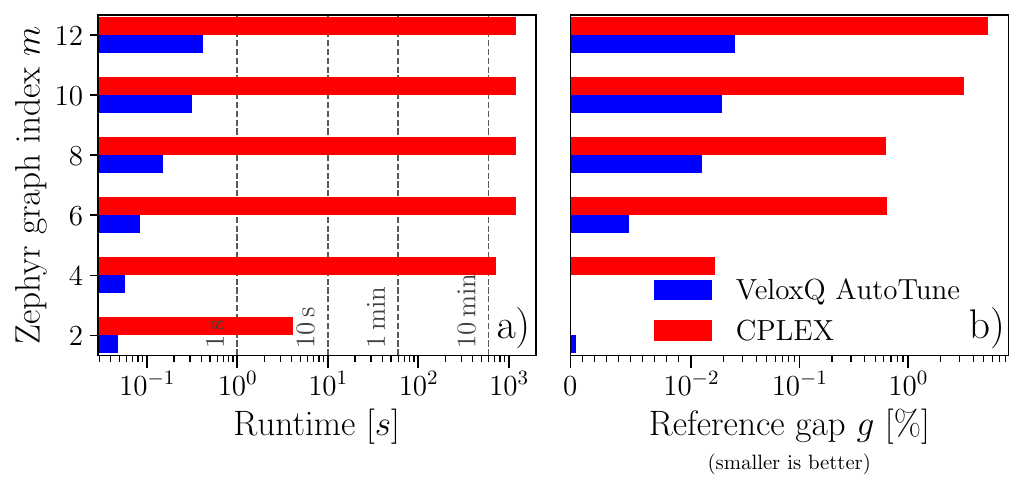}
    \caption{Comparison of VeloxQ with CPLEX on problems with Zephyr topology, analogous to the cases shown in Fig.~\ref{fig:zephyr_native}.
        Panel \textbf{a)} shows the average runtime, while panel \textbf{b)} shows the reference gap relative to the SA solution.
        Due to the increased connectivity of the Zephyr graph, CPLEX is able to compete
        with VeloxQ only for the smallest instance, \(Z_2\), and quickly exhausts the time limit of 1000s
        for larger instances.
    }
    \label{fig:zephyr_cplex}
\end{figure}

Finally, we consider the newest D-Wave platform, the Advantage2 QPU. As of
the beginning of 2026, the device topology corresponds to an imperfect Zephyr graph \(Z_{12}\)
with $4580$ working qubits, which limits the problem sizes for which we can conduct a 1-to-1 comparison with the QPU.
Motivated by the performance of VeloxQ on the Pegasus topology, we extend the benchmark far beyond the capabilities of
current quantum hardware, up to a \(Z_{1750}\) graph with more than $9.8 \times 10^7$ variables.
As an illustrative estimate based on historical D-Wave device releases, a quantum annealer capable of natively handling instances of this scale would not become available for roughly three decades under a simple extrapolation (cf. Fig.~\ref{fig:dwave_roadmap}).
This estimate assumes that the size of the current Advantage2 device would double every two years, which is consistent with previous D-Wave releases, and that
scaling challenges such as hardware reliability (e.g., the number of idle qubits), control electronics, and communication bandwidth would not become limiting factors.
Technical details of the computational experiments remain the same as in the previous case, with the
new results presented in Fig.~\ref{fig:zephyr_native}. Once again, VeloxQ is able to closely match
both the quantum annealer and the quantum-enabled hybrid solver in terms of solution quality and reported runtime on the benchmarked sizes.
A comparison with CPLEX on the same Zephyr instances is shown in Fig.~\ref{fig:zephyr_cplex}: due to the higher connectivity of the Zephyr graph, CPLEX is competitive only on \(Z_2\) and quickly exhausts the 1000\,s runtime limit for larger instances. These results further support
the scalability of VeloxQ and its competitiveness even in regimes far beyond the capabilities
of current Zephyr-based quantum hardware.

\subsection{Comparison of VeloxQ~SDK and D-Wave~Ocean~SDK}
\label{sec:dwave_sdk_comparison}

VeloxQ can be accessed through the Python package \texttt{veloxq\_sdk}~\cite{sdk}.
The SDK is designed to expose VeloxQ as a cloud backend while keeping the user-side
interface close to common QUBO and Ising workflows. Problem instances can be supplied as
in-memory biases and couplings, sparse dictionaries, NumPy arrays, \texttt{dimod} models,
or files. The following self-contained example creates a \texttt{dimod}
\texttt{BinaryQuadraticModel} and submits the same instance to VeloxQ and to a D-Wave
QPU, assuming the required API~tokens are available in the environment:

\begin{lstlisting}
import os
import dimod
from dwave.system import DWaveSampler
from dwave.system import EmbeddingComposite
from veloxq_sdk import VeloxQSolver

# Simple BQM example
linear = {0: 1.0, 2: -1.0}
quadratic = {
    (0, 1): -1.0,
    (1, 2): 0.5,
}
bqm = dimod.BinaryQuadraticModel(
    linear,
    quadratic,
    0.0,
    dimod.SPIN,
)

velox_solver = VeloxQSolver()
velox_result = velox_solver.sample(bqm)
print('VeloxQ:', velox_result.first)

qpu = DWaveSampler()
dwave_sampler = EmbeddingComposite(qpu)
dwave_result = dwave_sampler.sample(bqm)
print('D-Wave:', dwave_result.first)
\end{lstlisting}
Further examples are available in the \texttt{veloxq\_sdk} repository~\cite{sdk}.

When comparing cloud-accessible solvers, the runtime accounting becomes a subtle issue,
and care must be taken to ensure a fair comparison between different platforms. For D-Wave~QPUs,
the \texttt{qpu\_access\_time} reported by the API measures
the time for which the QPU is accessed by a submitted problem. It includes QPU programming
and the repeated anneal-read sampling cycles, including thermalization-related delays.
This quantity is not the same as the externally measurable wall time of a solver call
through the SDK. The wall time also includes API~communication, queuing and service overheads visible
to the client, submission/upload, result retrieval, and other remote-access overheads.
A common pitfall in D-Wave benchmarks is to use the annealing time of a single run
as a proxy for the total runtime, which can lead to a substantial underestimation of the user-experienced wall time.

For VeloxQ, we distinguish three time measurements. The solver time is the actual backend computation
time on GPU. The job time includes the solver computation and platform-side API~orchestration, but
excludes upload and download. The wall time is the full client-observed elapsed time,
including upload, platform/API~overhead, computation, and result download.
This distinction is especially important for short jobs, for which the externally observed
runtime may be dominated by API and data-transfer costs rather than by the solver-internal
computation.

\begin{figure}[!tbp]
    \centering
    \includegraphics[width=0.85\linewidth]{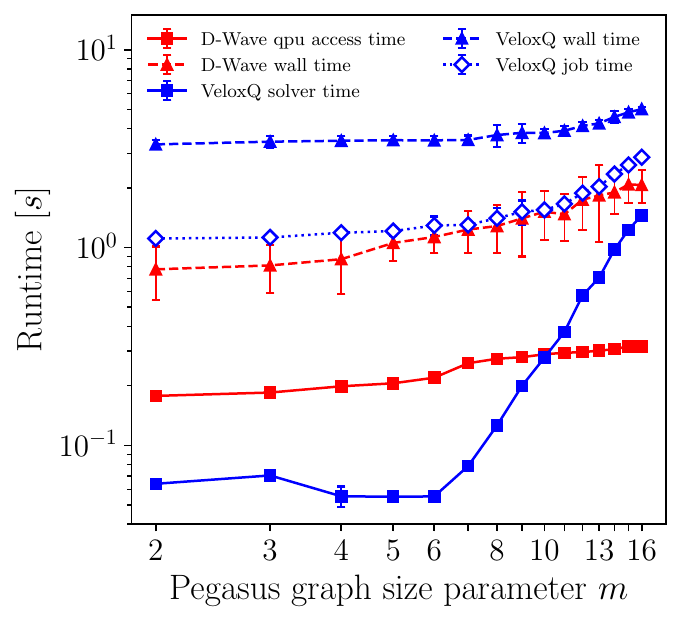}
    \caption{Runtime accounting for native Pegasus QUBO instances submitted through
        cloud SDKs. Red markers show D-Wave timings: \texttt{qpu\_access\_time},
        which includes QPU programming and anneal-read sampling cycles, and the
        externally measured wall time of the SDK call. Blue markers show VeloxQ
        time, job time, and wall time. Solver time measures backend computation
        only, while job time additionally includes platform-side orchestration but
        excludes upload and download. Finally, wall time is the full client-observed elapsed time.
        The comparison shows that API and data-transfer overheads are a substantial part
        of the user-visible runtime, especially for short solver executions (in small instance regime).}
    \label{fig:pegasus_runtime_sdk}
\end{figure}

In Fig.~\ref{fig:pegasus_runtime_sdk} we present the comparison of these time measurements
on the same set of native Pegasus instances as in Fig.~\ref{fig:pegasus_native}. In this regime,
both D-Wave and VeloxQ wall times are dominated by API and data-transfer overheads. However,
VeloxQ is able to handle natively much larger problems, for which one expects the solver time
to become dominant. In fact, the trend for $P_{m>13}$ instances suggests that the three
time measurements may converge for problems not much larger than the $P_{16}$ graph.

\section{Benchmarks against Kipu Quantum Solver}
\label{sec:benchmarks_hubo}

In this section we compare VeloxQ to the Kipu Quantum solver BF-DCQO (Bias-Field Digital Counterdiabatic Quantum Optimization)~\cite{romero2024}, which is a
digital-quantum algorithm for solving higher-order unconstrained binary optimization (HUBO) problems. QUBO solvers, such as VeloxQ or quantum annealers, can be used to
solve HUBO problems (such as 2-SAT) at the expense of an increased number of variables and higher problem complexity~\cite{dattani2019}. A key conceptual advantage of the Kipu Quantum algorithm over QUBO solvers is
that it solves HUBO problems natively and avoids the costly HUBO-to-QUBO reduction. In the benchmarks we compare
the performance of VeloxQ to results obtained with the Kipu Quantum solver BF-DCQO~\cite{romero2024}.
This benchmark separates two questions: whether VeloxQ remains competitive in solution quality after HUBO-to-QUBO reduction, and whether its QUBO scalability offsets the auxiliary-variable overhead.

\subsection{Higher-order unconstrained binary optimization}

The benchmark consists of two types of problems: NP-complete 3-Satisfiability (3-SAT) and one-dimensional random spin glass with three-body interactions. These problems
were selected to facilitate a direct comparison with the results of the Kipu Quantum solver BF-DCQO, which in part were obtained using digital quantum hardware,
the IBM Fez quantum platform. The chain structure of these problems is particularly well suited to the topology of existing IBM quantum processors~\cite{romero2024}.

\begin{figure}[!tbp]
    \centering
    \includegraphics[width=\linewidth]{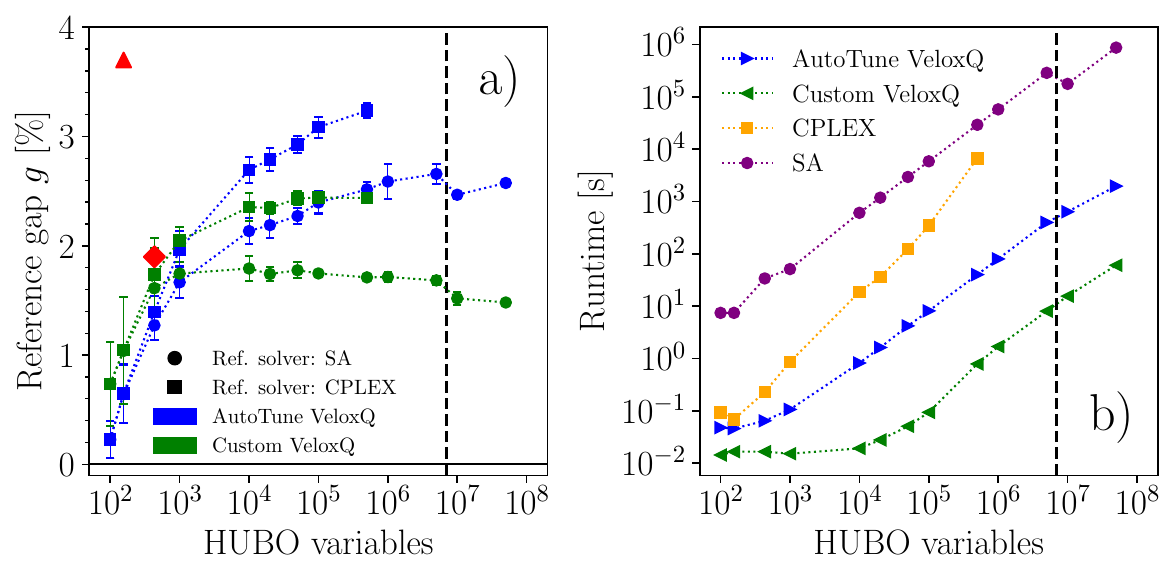}
    \caption{Results of the benchmarks for the random three-body Ising model, Eq.~\eqref{eq:nn}.
        \textbf{a)} Quality of the solutions obtained by VeloxQ, measured by the relative
        gap to the reference solution from CPLEX (squares) and Simulated Annealing (SA) (circles),
        with default (blue) and custom (green) settings. Red triangle and diamond
        denote BF-DCQO results on real and simulated quantum hardware, respectively,
        as reported in~\cite{romero2024}. \textbf{b)} Runtimes of VeloxQ for default (blue right triangles)
        and custom (green left triangles) settings, alongside the times needed to obtain the reference solutions
        with CPLEX (orange squares) and SA (purple circles). Results were averaged over 10 random instances for each problem size
        below $5\times 10^{6}$ variables, and over 2--5 instances for larger problems.
        For instances with more than $5\times 10^{6}$ variables (indicated with vertical dashed line),
        SA was run with modified parameters to keep the reference computations tractable.}
    \label{fig:3BodyIsing}
\end{figure}

Because test instances were not provided in Ref.~\cite{romero2024}, we generated our own benchmark set.
For HUBO instances corresponding to a one-dimensional spin glass, the cost function is equivalent to the Hamiltonian
\begin{align}
    H^{\text{NN}}(\vb{s}) & = \sum_{i=1}^{N} h_i s_i + \sum_{i=1}^{N-1} J_{i} s_i s_{i+1} + \sum_{i=1}^{N-2} K_{i} s_i s_{i+1} s_{i+2} \label{eq:nn},
\end{align}
where the couplings $h_i, J_i, K_{i}$ are drawn randomly from a Gaussian distribution with zero mean and unit variance.

The cost function corresponding to the weighted MAX 3-SAT problem is given by the Hamiltonian
\begin{align}
    H^{\text{MW3S}}(\vb{s}) & = \sum_{i=1}^{N-2} \frac{\omega_i}{8} \left[1 + (-1)^{c_i} s_i\right]
    \left[1 + (-1)^{c_{i+1}} s_{i+1}\right] \nonumber                                               \\
                            & \times \left[1 + (-1)^{c_{i+2}} s_{i+2}\right], \label{eq:mw3s}
\end{align}
and it aims to maximize the weighted sum of satisfied clauses,
where each clause consists of three propositional variables and is weighted by a random value \(\omega_i\), drawn from a uniform distribution
on the interval \([0,1]\). The \(c_i\) are either 0 or 1, chosen randomly, and determine whether the corresponding propositional variable is negated or not.

As VeloxQ can currently only handle second-order cost functions natively, we perform a reduction of the form
\begin{align}
    \pm s_i s_j s_k & \to 3 \pm \left(s_i + s_j + s_k + 2s^{\text{aux}}\right) \nonumber                                  \\
                    & + 2 s^{\text{aux}} \left(s_i + s_j + s_k\right) + s_i s_j + s_j s_k + s_i s_k, \label{eq:reduction}
\end{align}
where \(s^{\text{aux}}\) is an auxiliary variable~\cite{dattani2019}. This has a negative impact on the achievable problem sizes because it introduces one additional variable
for each third-order term, which can quickly become prohibitive for large problems. The results in the following sections show, however, that
VeloxQ has sufficient scalability to handle HUBO problems of considerable size after reduction.

\subsection{Benchmark results}

\begin{figure}[!tbp]
    \centering
    \includegraphics[width=\linewidth]{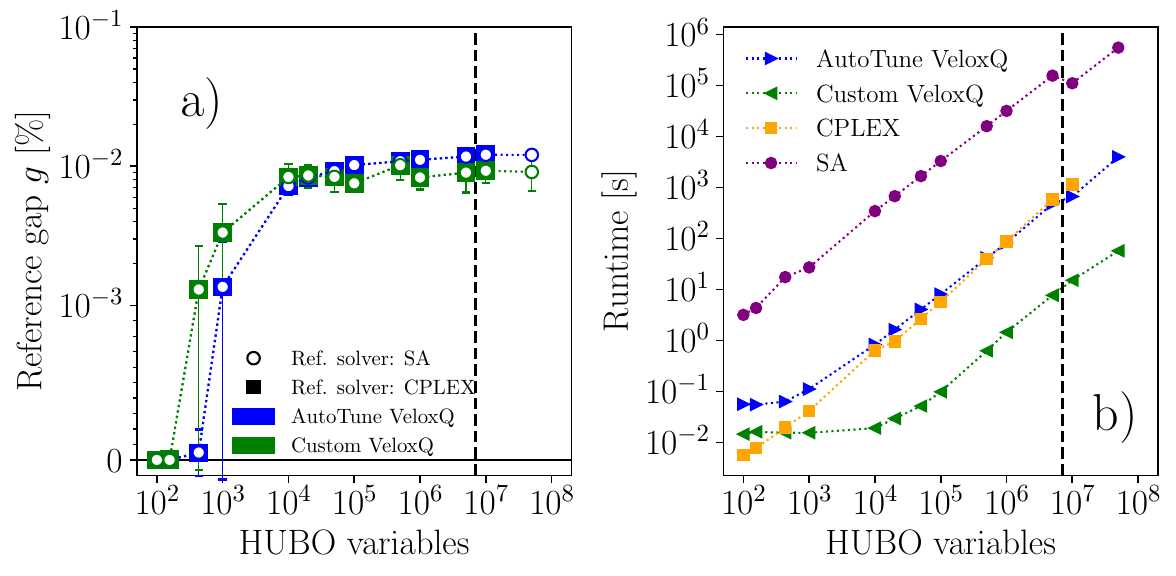}
    \caption{Results for the weighted MAX 3-SAT problem, given by Eq.~\eqref{eq:mw3s}. Contents of panels \textbf{a)} and \textbf{b)} are analogous to Fig.~\ref{fig:3BodyIsing}.}
    \label{fig:weighted_max3SAT}
\end{figure}

The left panels of Figs.~\ref{fig:3BodyIsing} and~\ref{fig:weighted_max3SAT} show the quality of the solutions obtained by VeloxQ,
measured by the relative distance to two types of reference solution: those obtained by the industry-standard CPLEX solver and by the SA algorithm.
The right panels display the corresponding times needed to obtain the solutions.

Across a broad range of problem sizes, from a modest \(100\) variables, up to $5 \times 10^{7}$ (
$\sim 10^{8}$ variables after HUBO \(\xrightarrow{}\) QUBO reduction), VeloxQ remains competitive, with reference gaps
in the \(2-3\)\% range in the structureless random case, and below \(0.01\)\% for the structured MAX 3-SAT problems. Comparison of runtimes
with CPLEX demonstrates another advantage of VeloxQ, namely a predictable and consistent
scaling with problem size, depending in principle only on the density of the problem and not any particular structure.
For the structured weighted MAX 3-SAT instances, CPLEX exploits the problem structure to reach high-quality solutions much faster than SA; nevertheless, VeloxQ stays within 0.01\% of the reference and is more than an order of magnitude faster for instances above \(10^4\) variables.
In the benchmark considered here, VeloxQ also improves on the reported BF-DCQO results~\cite{romero2024} in solution quality and tractable problem size, both in computer simulation (red diamonds) and in execution on a real IBM quantum platform (red triangles).
This comparison shows that, for these HUBO-derived instances, scalable QUBO solving can remain effective despite the auxiliary-variable overhead of reduction.

\section{Benchmarks against solvers with ground state certification}
\label{sec:benchmarks_certification}

A generic QUBO problem is NP-hard, which implies that one cannot expect to find the ground state in polynomial time~\cite{lu2010hybrid}.
Nevertheless, provably optimal solutions, even for relatively small problems, are very valuable, e.g., for assessing the performance
of heuristic solvers, classical or quantum. Beyond benchmarking, exact solvers can be useful as subroutines
in various hybrid algorithms, which decompose large scale problems into smaller, manageable chunks and then merge
the individual solutions to find a low-energy state of the original problem~\cite{Zintchenko2015}.
We can distinguish essentially two types of exact QUBO/Ising solvers: brute-force (BF), which
systematically explores the entire solution space, and specialized solvers that exploit the structure of a problem to
guarantee reaching a ground state. The former type is completely general, i.e., able to handle any QUBO instance regardless
of its structure, but due to exponential growth of the solution space, it is only feasible for small problems, up
to \(\sim 60\) variables with sophisticated implementation~\cite{BF} on modern hardware. In the latter case,
generality is traded for efficiency, as the solvers are tailored to specific problem classes, and in turn can handle
larger instances, sometimes even up to \(\sim 1000\) variables for sufficiently simple topologies~\cite{Liu2021}.

In this section we compare VeloxQ to solvers with ground-state certification. First, VeloxQ is benchmarked against the efficient implementation of the brute-force solver~\cite{BF}.
Then, as examples of solvers exploiting the problem structure, we take into consideration the BEIT solver~\cite{BEITQsolver,BEIT_AWS}, and a method based on tropical tensor networks (TTN)~\cite{Liu2021}.
The purpose of this section is to compare the computational cost of certification with the speed of an uncertified heuristic that can nevertheless reach the same energy on modest instances.

\subsection{Brute force solver}

\begin{figure}[!tbp]
    \centering
    \includegraphics[width=\linewidth]{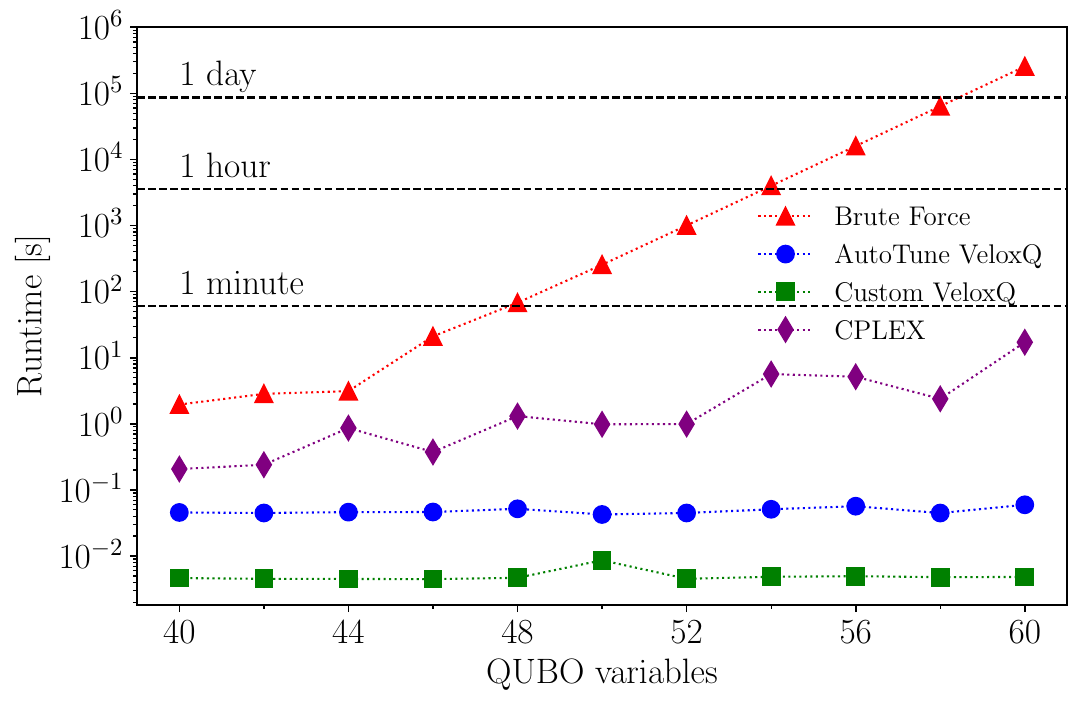}
    \caption{Benchmark results of VeloxQ against the state-of-the-art brute-force solver (BF). Red triangles denote the BF results, whereas green squares and blue
        circles represent VeloxQ in Custom and AutoTune modes, respectively. Purple diamonds denote results from CPLEX, which has also been able to find the exact ground states.
        Horizontal guidelines correspond to runtimes of one second, minute and hour.
        The runtime of the BF solver grows exponentially with the problem size, and for instances larger than 58 variables already exceeds the one-day mark.
        VeloxQ runs in constant time regardless of the variable count, and is able to match the quality of the BF solutions in a fraction of a~second,
        going as low as $\sim 5\,{\rm ms}$ with custom settings.
    }
    \label{fig:bf_runtime}
\end{figure}

To illustrate the ability of VeloxQ to produce exact solutions of small problems, we compare it against a~state-of-the-art, GPU-accelerated exact solver
based on a distributed brute-force approach~\cite{BF}. Due to the exponential growth of the solution space, we consider problems with up to $60$ variables with
all-to-all connectivity and random couplings from the range \([-1,1]\), which can be solved exactly in approximately three days using $2\times4$ NVIDIA H100 GPUs.
VeloxQ is able to match the quality of the solutions, i.e., obtain the same ground states, within a fraction of a~second, with runtime essentially constant across all considered instance sizes. Detailed runtime results are presented in Fig.~\ref{fig:bf_runtime}.

As for the larger problem sizes in Fig.~\ref{fig:pegasus_embedded}, we have already demonstrated that VeloxQ is capable of solving fully connected
random (structureless) problems with up to $160$ variables on similar timescales, obtaining configurations
of vanishing gap with respect to the ones obtained with the industry-standard CPLEX solver, taking up to $20$ minutes for the largest instances.
Note that the maximal size of the problem was limited by the maximal clique embedded into the Pegasus graph \(P_{16}\), and not by VeloxQ itself.
Even though VeloxQ does not offer a full certification of the ground state, the near-zero reference gaps indicate states that are either ground states or very close to them under the available CPLEX reference. At the same time, the considered problem sizes are well beyond the practical range of even the
best brute-force algorithms.

\subsection{Structure-exploiting solvers}

The aim of the second part is to compare VeloxQ against the other class of exact solvers, in particular two concrete examples:
an exact QUBO solver for the Chimera topology, offered as one of the commercial solutions by BEIT Inc.~\cite{BEITQsolver,BEIT_AWS}, and a specialized solver based on the
concept of TTN, introduced by~\cite{Liu2021}.
Out of the three considered options, the proprietary BEIT solver, utilizing dynamic programming on bounded treewidth graphs,
is the most limited in terms of the problem structure, as it can only handle
problems with up to \(1024\) qubits, arranged in a \(8\cross 16\) Chimera lattice. Furthermore, the QUBO couplings are restricted to
integers in the range \([-31, 31]\). By contrast, the TTN solver is more general, as it has no inherent limitations on the problem structure.
It exploits the tensor network representation of the Ising Hamiltonian, and an observation that the zero-temperature limit of the partition function
(which encodes the ground state energy) finds a natural formulation in terms of the so-called tropical algebra, which is a semiring over
\(\mathbb{R}\cup\{-\infty\}\) with addition replaced by the maximum operation, \( x \oplus y = \max(x,y)\), and multiplication by the sum,
\(x \otimes y = x + y\). To obtain the corresponding state, automatic differentiation of the contraction outcome with respect to the
tensor elements is performed. However, the complexity of tensor network contraction depends on the structure of the graph underlying the QUBO problem,
and so the TTN solver performs best on relatively sparse problems.

\begin{figure}[!tbp]
    \centering
    \includegraphics[width=\linewidth]{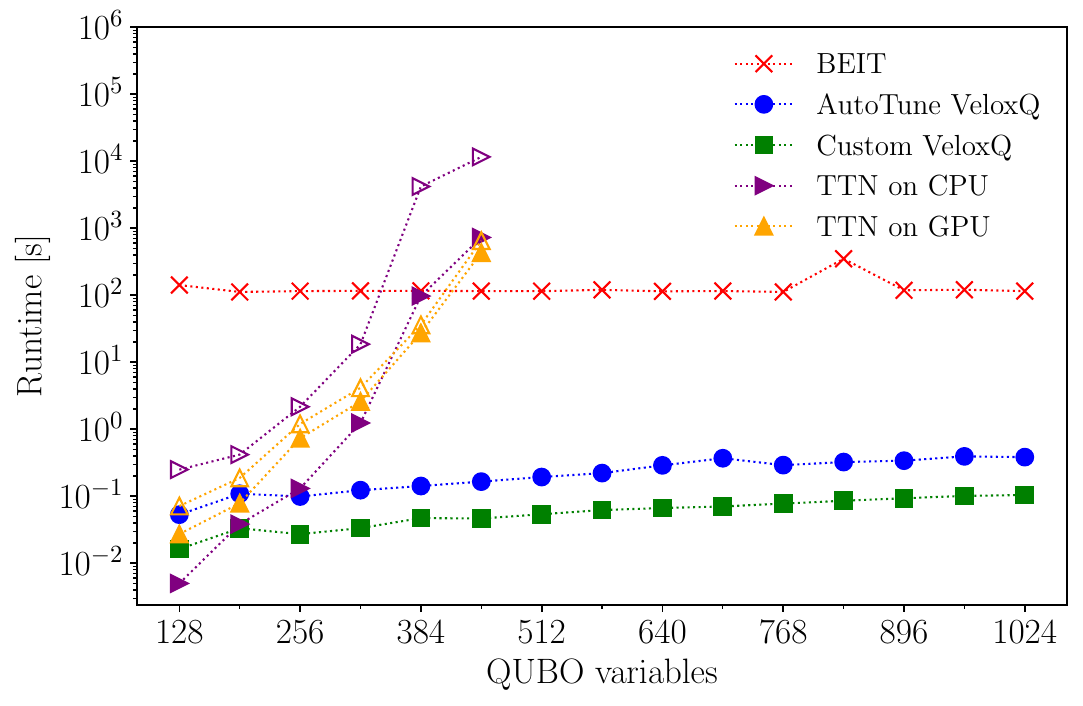}
    \caption{Runtime benchmark of VeloxQ against solvers with ground-state certification; runtimes only are shown because all solvers found optimal solutions. Red crosses: BEIT solver (runtime obscured by AWS-related overheads). Blue circles and green squares: VeloxQ in AutoTune and Custom modes, respectively. Filled triangles: TTN in energy-only mode. Empty triangles: TTN with state retrieval (differentiating through the network).}
    \label{fig:gs_certif}
\end{figure}

To assert the fairness of the benchmark, we investigate the performance of these three solvers on a most general problem that can be handled by all of them,
namely random Chimera instances, with sizes ranging from $128$ (\(C_{2,8}\)) to $1024$ (\(C_{8,16}\)) variables, and random integer couplings in the range \([-31,31]\).
We consider one random instance per graph size.
Importantly, all solvers obtained the lowest energy solution for instances they were able to process,
and so the comparison is based on the time needed to reach the solution. The results are presented in Fig.~\ref{fig:gs_certif}.
The BEIT Chimera solver is available only through the AWS cloud service, which makes the true runtime difficult to disentangle from the request overhead~\cite{BEITQsolver,BEIT_AWS}.
This is visible in the nearly constant time to solution across instance sizes, so the actual computation time is likely much shorter.
Regardless, an effective runtime of approximately 2 minutes for a certified ground state on 1024 variables remains a notable result.

Both the BEIT solver and VeloxQ always return the state alongside its energy.
The TTN solver, by contrast, can be run in two modes: computing only the lowest energy, or returning both the energy and the corresponding state.
On CPU, retrieving the state is significantly more time-consuming, increasing the runtime by up to 3 orders of magnitude; the GPU version does not suffer from such a slowdown.
Nevertheless, the complexity of tensor contraction increases rapidly with problem size, and the TTN solver cannot handle instances with more than $448$ variables, corresponding to the \(C_{8,7}\) Chimera graph.

VeloxQ delivered optimal configurations in time much shorter than other solvers, except for the two smallest instances.
For the energy-only TTN mode, contraction is faster on CPU than on GPU; the gap effectively disappears when the state itself is retrieved.
Although VeloxQ does not certify ground states, on these modest-size instances it matches the solution quality of the specialized solvers without sacrificing generality or scalability.
These results show that, when a ground-state certificate is not a strict requirement, VeloxQ can be a competitive alternative to both open-source and proprietary specialized solutions on modest-size instances.

\section{Benchmarks against physics-inspired algorithms}
\label{sec:physicsInpiredAlgorithms}

Ever since the introduction of the SA algorithm in the 1980s~\cite{SA}, physical processes have
been a rich source of inspiration for optimization algorithms. The idea is to map the optimization problem onto a physical system in such a way
that the optimal solution corresponds to the lowest energy state, the so-called ground state, of the system. The system is then evolved according to the laws of physics,
which, under appropriate conditions, favor the evolution towards low-energy states, akin to a ball rolling down a hill. In this section, we compare VeloxQ against two such state-of-the-art algorithms:
Parallel Annealing (PA) and the Simulated Bifurcation Machine (SBM), both rooted in quantum adiabatic optimization.
This benchmark uses planted-solution families because they provide known ground-state energies without relying on a separate reference solver.

PA is a variant of an annealing algorithm, designed to leverage the massively parallel nature of modern hardware accelerators~\cite{Jiang2023}. As this is a classical algorithm,
quantum spins are first replaced by classical spins, i.e., binary variables taking values \(\pm 1\). These spins are subsequently relaxed to analog variables in the range \([-1,1]\),
emulating the idea of quantum superposition. Real classical spins are recovered via the sign function. The initial Hamiltonian is taken to be a
convex function with an easy-to-find global minimum (ground state), such as \( x^2 \), and gradually
shifted towards the target Hamiltonian, encoding the QUBO problem in Ising form. Since this is a~simulation of a physical process, the analog spins do not update automatically, and
a suitable update rule must be chosen. In the case of PA, it is a modified version of gradient descent called the straight-through estimator. For the purposes of this benchmark,
we have created a custom, GPU-accelerated implementation of the PA algorithm. Technical details about PA are outlined in Appendix~\ref{app:solvers}.

SBM, an Ising solver implementing the idea of simulated bifurcation, was first proposed in 2016~\cite{Goto2016} by researchers at Toshiba,
in the form of a quantum computer consisting of a nonlinear oscillator network, adiabatically driven through a bifurcation point,
and generating a superposition of quantum states that encodes the solution to a given combinatorial optimization problem. Since appropriate quantum hardware is not yet readily available,
and the simulation of a system working as a general-purpose quantum computer is computationally infeasible, subsequent work proposed a sequence of approximations, yielding
a classical nonlinear dynamical system, governed by non-autonomous Hamiltonian equations of motion for harmonic oscillators with quartic nonlinearity,
coupled via interaction that encodes the Ising problem~\cite{Goto2019}. The equation of an individual oscillator is also known under the name Duffing equation.
This quantum-inspired solver leverages \emph{classical} chaos and bifurcation to explore the solution space.
Initially dominated by the parabolic minimum, post-bifurcation the energy landscape approximately encodes the local minima of the Ising term, and the system flows toward low-energy solutions; binary states are then extracted as the sign of the analog variables.
The chaotic dynamics also allow natural parallelization: many initial states can be evolved simultaneously and the best solution selected.

A commercial optimization suite based on a modified version of the SBM algorithm~\cite{GotoSBM}, called SQBM+, is available from Toshiba Digital Solutions Corporation.
Due to restricted solver access, we could not compare SQBM+ directly with VeloxQ.
Instead, we replicated the original SBM algorithm with GPU acceleration and used it as a reference in the benchmarks.
Technical details of SBM are summarized in Appendix~\ref{app:solvers}.

Since both our physics-inspired solvers of choice and VeloxQ are topology-agnostic, we benchmark them on three families of random instances with planted solutions:
\begin{itemize}
    \item \textbf{3-regular 3-XORSAT (3R3X)}~\cite{Hen2019} --- tests performance on instances with a golf-course-like landscape and low-lying excited states at extensive Hamming distance from the planted solution.
    \item \textbf{Tile planting (TP)}~\cite{Perera2020} --- tests robustness to tunable frustration in sparse lattice instances.
    \item \textbf{Wishart planting (WP)}~\cite{Hamze2020} --- tests dense zero-field Ising instances with tunable ruggedness, generated from random-constraint constructions with a planted ground state.
\end{itemize}
This setup yields problems with exactly known ground-state energies, allowing us to compare solvers without a separate reference solver.
TP and WP instances are particularly informative because their \`hardness' is continuously tunable; by \`hardness' we mean empirically observed typical difficulty, not the more precise computer-science notion of computational complexity.
To streamline the preparation of problems, we employ an open-source Python-based tool called \emph{Chook}~\cite{Perera2021}.

\subsection{3-regular 3-XORSAT equation planting}

The starting point for the generation of 3-regular 3-XORSAT (3R3X) instances is a system of linear equations over \(\mathbb{Z}_2\),
\begin{equation}
    \sum_{j=1}^{n} a_{ij} x_j = b_i \mod 2, \quad \text{for} \quad i = 1, \ldots, m,
\end{equation}
where \(x_j \in \{0,1\}\) are binary variables. Specifically, we consider a case wherein each equation involves exactly three randomly chosen variables,
and each variable appears in exactly three equations --- hence the name 3-regular 3-XORSAT. Crucially, this system of equations can be solved in polynomial time,
using e.g., Gaussian elimination. Its solution will be the planted ground state of the corresponding optimization problem.
Translated into Ising form, these equations can be written as
\begin{equation}
    \prod_{j: a_{ij} = 1} s_j = (-1)^{b_i}, \quad \text{for} \quad i = 1, \ldots, m,
\end{equation}
where \(s_j \in \{-1,1\}\) are the Ising spins. By squaring and adding the equations, we can obtain the relevant Ising Hamiltonian
\begin{align}
    H_{\text{3R3X}} = & \sum_{i=1}^{m} \left((-1)^{b_i} - \prod_{j: a_{ij} = 1} s_j \right)^2 \nonumber \\
    \triangleq        & - \sum_{i=1}^{m} (-1)^{b_i} \prod_{j: a_{ij} = 1} s_j \nonumber                 \\
    =                 & - \sum_{i=1}^{m} J_i s_{i_1} s_{i_2} s_{i_3},
\end{align}
where \(\triangleq\) denotes equality up to an irrelevant constant term. For each equation \(i\) there are exactly three indices \(j \in \{i_1, i_2, i_3\}\) such that \(a_{ij} = 1\), and the couplings \(J_i = (-1)^{b_i}\) encode the right-hand sides of the equations.
Thus, the cost function for this problem is composed of \(m\) three-body Ising terms, which can be reduced to the usual two-body variant with the reduction scheme given by Eq.~\eqref{eq:reduction},
yielding finally a \(2m\) variable QUBO/Ising problem. Obtaining an optimal solution of such a problem is a challenging task for QUBO solvers, because the low-lying excited states are
far away (in the sense of Hamming distance) from the ground state, and so it is easy to get stuck in a local minimum~\cite{Hen2019}.

\begin{figure}[!tbp]
    \centering
    \includegraphics[width=\linewidth]{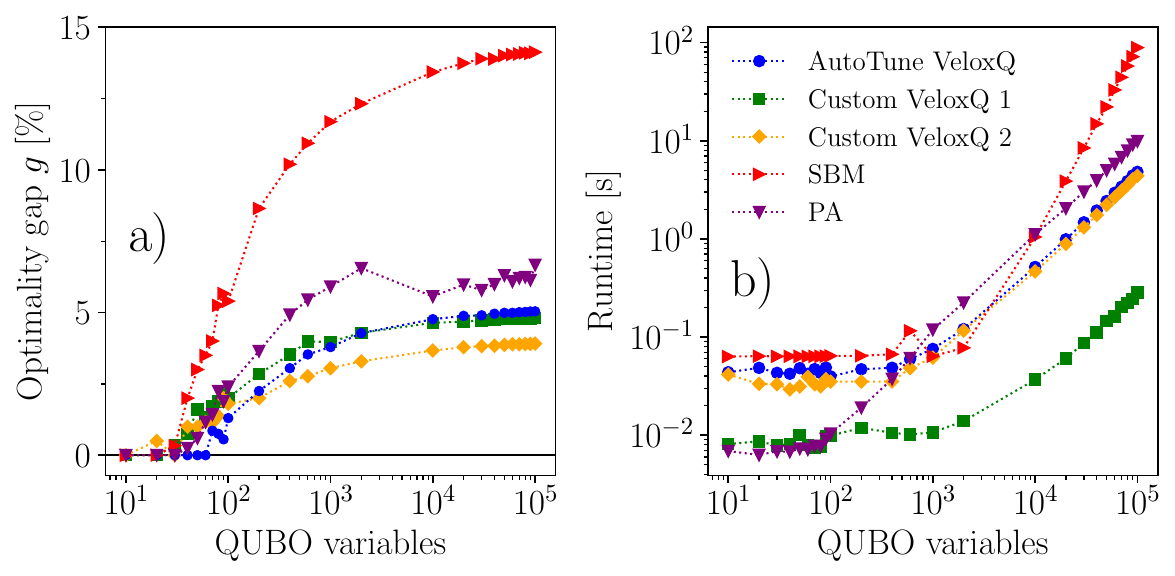}
    \caption{Results of the benchmarks of VeloxQ against the PA and SBM solvers on 3R3X instances, \textbf{a)} optimality gap and \textbf{b)} runtime in seconds.
        We show data points for VeloxQ in AutoTune mode (blue circles) and two manually optimized versions, for runtime (green squares) and
        for quality of the solution (orange diamonds). For problems with up to \(50\) variables, the optimization
        procedure consistently returns a ground-state solution. For larger instances, the optimality gap increases but does not exceed \(5\%\).
        High-quality solutions are also delivered by parallel annealing (purple inverted triangles), with a runtime that differs only by a constant factor from VeloxQ in the default version.
        The SBM (red right-facing triangles) is less competitive on this benchmark, with optimality gaps growing rapidly with system size, finally reaching almost \(15\%\)
        and runtimes demonstrating worse scaling in the 1000+ variable regime. Moreover, this result was obtained after extensive tuning of the SBM parameters, which should be compared
        with the solid out-of-the-box performance of VeloxQ (AutoTune mode).
    }
    \label{fig:3R3X}
\end{figure}

Using the procedure outlined above, we prepared a set of random 3R3X instances with sizes ranging from 10 to $10^{5}$ variables, with 10 instances per size. In Fig.~\ref{fig:3R3X}
we present the averaged results of optimality gap and runtimes. The error bars are omitted for the sake of clarity.
The outcome of this computational experiment demonstrates that VeloxQ is competitive with modern, physics-inspired solvers for this planted-instance family.
For problems with up to \(50\) variables, the optimization procedure consistently returns a ground-state solution; for larger instances, the optimality gap increases but does not exceed \(5\%\).
PA also delivers high-quality solutions, with runtime differing from default VeloxQ by only a constant factor.
SBM is less competitive on this benchmark, even after tuning, with gaps increasing rapidly with system size.
The two custom VeloxQ settings show the expected trade-off between solution quality and runtime.

\subsection{Tile planting}

Tile planting (TP) instances are a type of square-lattice Ising model with periodic boundary conditions. They are constructed by partitioning the lattice graph \(G=(V,E)\) into unit-cell
tiles \(C \in \mathcal{C}\), creating a checkerboard pattern. Each tile \(C\) consists of four vertices, and each vertex is shared by two tiles. For each subgraph, the Hamiltonian is defined as
\begin{equation}
    H^{C}_{\text{tiled}} = - \sum_{\langle i,j \rangle \in E[C]} J_{ij} s_i s_j,
\end{equation}
where the sum runs over all edges in the tile. The full Hamiltonian is then given by the sum over all tiles
\begin{equation}
    H_{\text{tiled}} = \sum_{C \in \mathcal{C}} H^{C}_{\text{tiled}}.
\end{equation}
Importantly, if all the tile Hamiltonians have a common ground state, then the full Hamiltonian will also share the same ground state configuration.
This fact allows one to construct instances with a known minimum energy value~\cite{Perera2020}. The remaining freedom in choosing the coupling constants \(J_{ij}\) allows
one to tune the level of frustration in the system, and thus the hardness of the problem. Without loss of generality (due to gauge freedom), we can assume
that our target ground state is a fully polarized, ferromagnetic state \(\vb{t} = (+1,+1,\ldots, +1)\). Then, four types of nonequivalent tiles are introduced, \(C_i\) for \(i = 1, \ldots, 4\), such that
tile of type \(i\) has \(i\) ground states, always including the fully polarized ferromagnetic state. A complete TP instance is obtained by randomly selecting
the tile type for each tile, according to a given probability distribution \((p_1, p_2, p_3, p_4)\). It was shown that this three-dimensional space of control parameters
allows for a fine-tuning of the problem hardness~\cite{Perera2020}. For the purpose of our benchmark, we choose a particular 1-dimensional subspace in this parameter space,
given by \( S = \{(0, p_2, 0, 1-p_2) \mid p_2 \in [0,1]\}\), as it is the most versatile one-dimensional subspace out of those exhibiting an easy-hard transition. In general, there
are three such subspaces, since it is the fraction of \(C_2\) tiles that control the problem hardness.
\begin{figure}[!tbp]
    \centering
    \includegraphics[width=\linewidth]{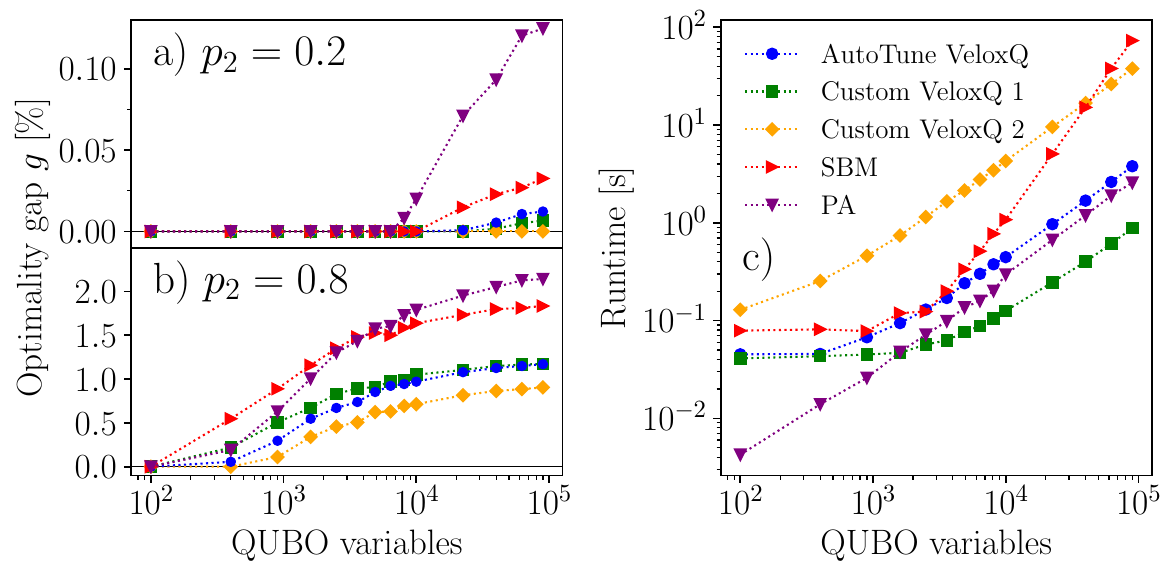}
    \caption{Results of the benchmarks of VeloxQ against the PA and SBM solvers on TP instances, optimality gap in the \textbf{a)} `easy' regime \((p_2 = 0.2)\) and
        \textbf{b)} `hard' \((p_2 = 0.8)\) regime. Panel \textbf{c)} shows the runtimes only in the \((p_2 = 0.8)\) case, since the value of \(p_2\) does not affect the runtime.
        We show data points for VeloxQ in AutoTune mode (blue circles) and two manually optimized versions, for runtime (green squares) and
        for quality of the solution (orange diamonds). Quality-optimized VeloxQ is the only tested solver able to obtain ground states for all sizes in the easy regime, and consistently
        outperforms other solvers in the hard regime. If runtime is the main concern, it is possible to decrease it by a factor of 3, while still
        maintaining the optimality gap below \(1\%\) in the \(p_2=0.8\) case. Parallel annealing (purple inverted triangles) behaves similarly to VeloxQ in most of the easy
        regime \((N\leq 10^4)\), but it produces noticeably worse (although still of high quality) solutions beyond it. Nevertheless, it has an advantage
        for problem sizes below \(10^3\) variables, since it is able to deliver solutions particularly fast.
        After appropriate hyperparameter tuning, the SBM (red right-facing triangles) in its original formulation demonstrates competitive performance, emerging
        on top of PA in the easy regime, and on par with it in the hard regime. VeloxQ is able to outperform it in both cases.
    }
    \label{fig:tile}
\end{figure}

Results for two kinds of TP instances from \(C_2-C_4\) subspace, easy \((p_2 = 0.2)\) and hard \((p_2=0.8)\), on square lattices with side of length
\(L \in [10, 300]\), are presented in Fig.~\ref{fig:tile}.
Since these instances are also randomized, for each size we prepared \(10\) copies and averaged the results, again omitting the error bars for clarity.
With moderate manual tuning, VeloxQ can consistently find the ground state for even the largest instances with 90k variables in the easy regime.
The runtime-optimized setting reduces runtime while keeping the hard-regime gap below \(1\%\).
PA remains competitive on smaller easy-regime instances, with larger gaps appearing at larger sizes, while tuned SBM is competitive with PA but does not exceed the best VeloxQ settings.
The default VeloxQ settings are already sufficient to outperform PA and SBM in both difficulty regimes.

\subsection{Wishart ensemble}
Both 3R3X and TP problems are formulated as Ising models with rather sparse coupling matrices. On the other
hand, Wishart ensemble planting (WP) provides a way to construct instances with known ground state
energy and all possible pairwise couplings. We begin by choosing a ground state configuration, which again can be chosen as the
fully polarized ferromagnetic state \(\vb{t}\), and subsequently concealed by gauge randomization.
Let now \(W\) be an \(N \times M\) real matrix, such that
\begin{equation}
    W^T \vb{t} = 0,
\end{equation}
i.e., describing a set of \(M\) linear equations in \(N\) variables, with the ground state \(\vb{t}\) as the solution. The
ratio \(\alpha = M/N\) of equations to variables determines the hardness of the problem.
This property guarantees that an Ising model defined as
\begin{equation}
    H_{\text{WP}} = - \frac{1}{2} \sum_{i,j} J_{ij} s_i s_j,
\end{equation}
with
\begin{align}
    \tilde{J} & = -\frac{1}{N} W W^T                  \\
    J         & = \tilde{J} - \text{diag}(\tilde{J}),
\end{align}
has a ground state \(\vb{s} = \vb{t}\) with energy
\begin{equation}
    E_0 = -\frac{1}{2} \vb{t}^T J \vb{t} = \frac{1}{2} \mathrm{Tr}(\tilde{J}).
\end{equation}
The actual construction of the matrix \(W\) is carried out by sampling each column vector \(\vb{w}\)
from a multivariate Gaussian distribution with zero mean, and covariance matrix \(\Sigma\),
given by
\begin{equation}
    \Sigma = \frac{N}{N-1}\left[\mathbb{1}_N - \frac{1}{N} \vb{t}\vb{t}^T\right],
\end{equation}
where \(\vb{t}\) denotes our planted ground state. It can be shown that \(\vb{w}^T \vb{t} = 0\) for all columns of \(W\),
which concludes the construction. The resulting random matrix \(W W^T\) follows a matrix
version of \(\chi^2\) distribution, known as the Wishart distribution~\cite{Hamze2020}.

\begin{figure}[!tbp]
    \centering
    \includegraphics[width=\linewidth]{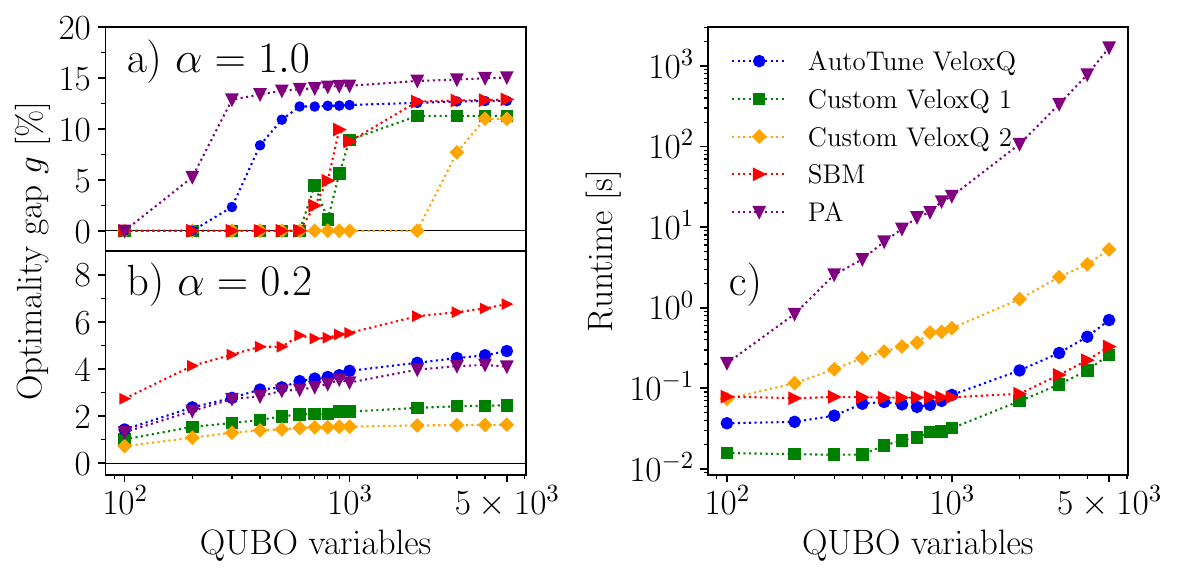}
    \caption{Results of the benchmarks of VeloxQ against PA and SBM on WP instances. Optimality gap in the \textbf{a)} ``easy'' regime \((\alpha = 1.0)\) and \textbf{b)} ``hard'' regime \((\alpha = 0.2)\). As in the TP case, panel \textbf{c)} shows runtimes only for \(\alpha = 0.2\), since the value of \(\alpha\) does not affect the runtime of the considered solvers. AutoTune VeloxQ: blue circles. Custom (runtime-tuned): green squares. Custom (quality-tuned): orange diamonds. PA: purple inverted triangles. SBM: red right-facing triangles.}
    \label{fig:wishart}
\end{figure}

When the equation-to-variable ratio \(\alpha \leq 1\), there exists a stable set of paramagnetic
configurations (uncorrelated with planted ground state), which essentially behaves as
a local minimum trap for classical heuristic solvers.
The generated problem is likely to be solved with high probability only if
the initialization happens by chance within the ground-state basin of attraction
(assuming no escape mechanism is present).
The parameter \(\alpha\) influences the size of this basin, with higher values
increasing the chances of a successful solution due to a fortunate initialization.
Therefore, in Fig.~\ref{fig:wishart} we present the benchmark results for two regimes, \(\alpha = 0.2\) (hard)
and \(\alpha = 1.0\) (easy). Since the WP instances are fully connected, the number of variables
is limited to \(N \leq 5000\). For each of the presented system sizes, ten instances were generated and the results averaged.
An interesting observation regarding the nature of `hardness' in the Wishart ensemble can
be made on the basis of the results presented in Fig.~\ref{fig:wishart}. In the easy regime, each solver
produces a ground-state solution up to a certain solver-dependent system size, after
which the solution quality rapidly degrades, finally settling at the optimality gap of around \(10-15\%\).
VeloxQ can be tuned to produce optimal configurations for
the broadest range of system sizes (up to \(N=1000\)), and the best results beyond this range. The hard regime
is characterized by a qualitatively different behavior, with all considered solvers
not reaching ground-state solutions under the tested settings even for the smallest instances with \(N=100\).
Even though the optimality gap increases with system size, the change is not as drastic as in the easy regime,
and VeloxQ is able to maintain the gap below \(5\%\) for the default setting and at the level
of \(1\%\) for the quality-optimized version.
PA underperforms in the easy regime but matches the default VeloxQ in the hard regime, at the cost of higher runtime; SBM is fast and accurate in the easy regime but degrades in the hard regime.

\section{Benchmarks against modern B\&B algorithms}
\label{sec:benchmarks_BB}

Branch-and-Bound (B\&B)~\cite{Wolsey1998} is a foundational framework for solving combinatorial optimization problems, including QUBO and Ising model formulations.
Over the years, numerous enhancements have been developed to improve its efficiency, focusing on solution-space pruning techniques and accelerating convergence.
In this section, we compare VeloxQ against selected B\&B algorithms, including a modern refinement of the algorithm developed by Quantagonia~\cite{Quantagonia2023}.
This benchmark focuses on dense pseudo-random instances where the search tree becomes expensive and pruning quality directly affects runtime and energy gap.

This comparison highlights differences in performance and scalability among the tested approaches.
Key evaluation metrics include solution quality, computational cost, and scalability to larger problem instances.
The selected B\&B methods construct solutions iteratively by solving incremental subproblems,
progressively appending spins to build the solution. This approach aligns naturally with the binary decision tree representation over fixed $\{-1, 1\}$ spin-like values.
Memory efficiency is enhanced by maintaining a pool of stored partial solutions, while subtree selection is guided by heuristic bound functions.
These functions are critical components that determine the effectiveness of each method. The choice of bound functions allows for a diverse range of methods~\cite{Hner2024},
including modern implementations adopted in the industry~\cite{Quantagonia2023, Chalkis2023, Huo2024}.

As reference methods for comparison to VeloxQ, we implemented search algorithms based on two bound functions. A standard approach to B\&B algorithms is represented by
$B_{\text{base}}$~\cite{Kochenberger2014}, in which the prefix subproblem energy is directly used as the bounding premise. The modern refinement of B\&B is denoted as $B_{\text{PSD}}$, in which
the bound relates to the subproblems transformed to positive semidefinite matrix formulations. This allows estimating energy minima for subproblems.
Such an approach, reflecting selected state-of-the-art techniques and modern advancements in optimization, aligns with the algorithmic outline described in~\cite{Quantagonia2023}.
PSD-based methods are highly adaptable, including successful applications to solving QUBO and Ising model problems in various computational environments, including quantum computing
\cite{Chalkis2023, Brand2022}. Our implementation, optimized for the CUDA runtime environment to facilitate a practical comparison with VeloxQ, is detailed in Appendix~\ref{app:BB}.

\begin{figure}[!tbp]
    \centering
    \includegraphics[width=\linewidth]{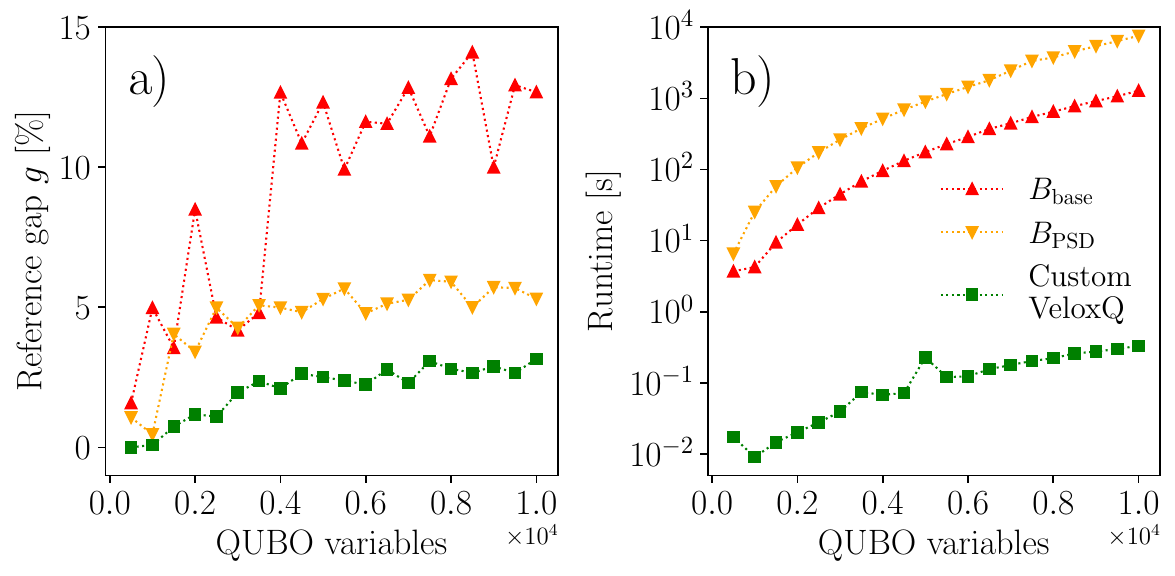}
    \caption{B\&B methods on dense, pseudo-random instances with up to $10\,000$ variables: \textbf{a)} reference gap, \textbf{b)} runtime. VeloxQ values were obtained using Custom VeloxQ with $150$ internal states. Reference energies were approximated by invoking VeloxQ iteratively, starting with $1000$ states and increasing by $50\%$ until the best energy value was repeated three consecutive times.}
    \label{fig:BBsz}
\end{figure}

The computation time and solution quality achieved by B\&B methods using $B_{\text{base}}$ and $B_{\text{PSD}}$,
in comparison to VeloxQ, are presented in Fig.~\ref{fig:BBsz}. To~ensure a diverse range of matrix sizes for benchmarking,
pseudo-random dense matrices were generated. While B\&B methods can operate on sparse matrices, the computational overhead is primarily
determined by matrix size, which corresponds to the height of the search tree. As such, dense matrices were chosen as the most suitable for this study.
Matrix sizes were distributed within the range of $1000$ to $10\,000$. Problem instances with smaller matrices were excluded, as they do not reflect the scope of the
method and are often more efficiently solved by brute-force approaches. Sizes up to $10^4$ were chosen as a practical upper limit for our computational setup, with individual runs requiring less than three hours.

The benchmark results show that VeloxQ obtains comparable or better solution quality than both B\&B algorithms. The most pronounced difference concerns the runtime: even for the smallest instances VeloxQ is two orders of magnitude faster than the
B\&B algorithms, and this difference increases with the growing instance size.
For instances exceeding $4\,000$ variables, the $B_{\text{base}}$ variant became unreliable, whereas the $B_{\text{PSD}}$ variant maintained more stable gaps at the cost of approximately $10\times$ longer runtime than $B_{\text{base}}$.

\section{Benchmark overview, limitations, and summary}
\label{sec:summary}

\subsection{Benchmark overview and limitations}
\label{sec:limitations_reproducibility}

\begin{table*}[t]
    \caption{Summary of the benchmark regimes considered in this paper.}
    \label{tab:benchmark_summary}
    \scriptsize
    \begin{ruledtabular}
        \begin{tabular}{p{0.13\textwidth}p{0.13\textwidth}p{0.17\textwidth}p{0.11\textwidth}p{0.15\textwidth}p{0.24\textwidth}}
            Benchmark         & Competitors                             & Instance type                                            & Size range                                           & Reference method             & Main outcome                                                                     \\
            \hline
            D-Wave native     & Advantage, Advantage2, Kerberos         & Pegasus and Zephyr sparse graphs                         & Up to \(P_{150}\), \(Z_{1750}\)                      & SA for non-certified regimes & Comparable quality on native sizes; VeloxQ extends to larger instances than QPU and Kerberos runs considered here \\
            D-Wave embedded   & Advantage                               & Dense complete graphs embedded on Pegasus                & Up to 160 logical variables                          & CPLEX                        & VeloxQ avoids embedding overhead and obtains low reference gaps                  \\
            HUBO              & BF-DCQO, CPLEX, SA                      & 3-body Ising and weighted MAX-3-SAT after QUBO reduction & Up to \(\sim 10^8\) QUBO variables            & CPLEX and SA where available & VeloxQ remains effective after reduction and reaches much larger instances       \\
            Certified solvers & BF, BEIT, TTN                           & Dense small instances and Chimera instances              & Up to 1024 variables for Chimera                     & Certified ground state       & VeloxQ often reaches the certified energy faster                                 \\
            Planted solutions & PA, SBM                                 & 3R3X, tile planting, Wishart                             & Up to \(10^5\) variables for sparse planted families & Planted ground-state energy  & VeloxQ is competitive across sparse and dense planted families                   \\
            B\&B              & \(B_{\text{base}}\), \(B_{\text{PSD}}\) & Dense pseudo-random instances                            & 1000--10000 variables                                & Iterative VeloxQ reference   & VeloxQ obtains comparable or better gaps with shorter runtimes in the tested range                 \\
        \end{tabular}
    \end{ruledtabular}
\end{table*}

Tab.~\ref{tab:benchmark_summary} provides a synoptic view of the six benchmark regimes considered in this paper, along with the competitors, instance types, size ranges, reference methods, and main outcomes.

Several aspects of the benchmark design should be kept in mind when interpreting the benchmark results.
First, VeloxQ is a heuristic solver and does not certify optimality except in cases where its output can be compared with an external certificate or planted solution.
Second, some reference gaps are measured relative to SA, CPLEX, or iterative VeloxQ references rather than known ground states.
Third, for the largest Zephyr instances, the paper demonstrates that VeloxQ can return samples at scale, but it does not claim independently certified solution quality when no reference solution is available.
Fourth, the tuning effort differs across solvers: VeloxQ is reported both in default and custom settings, while competitor settings follow either the cited study, default or recommended package settings, or the setup described in the relevant section.
Reproducibility details, including the role of the accompanying repository and API~access, are described in Sec.~\ref{sec:benchmark_methodology}.

\subsection{Summary}
This paper benchmarks VeloxQ across regimes chosen to test three aspects of QUBO-solver performance: solution quality, runtime, and scalability.
The comparison includes quantum annealers, a gate-based digital-quantum HUBO solver, physics-inspired algorithms, exact or certifying approaches, and modern Branch-and-Bound methods.
The benchmark instances were selected to include cases that are favorable to the comparison methods, such as native quantum-annealer topologies, HUBO instances close to the Kipu Quantum setting, planted-solution families, Chimera instances for BEIT, and dense instances for B\&B.
This design reduces the risk that the comparison is driven only by unfavorable instance choices for the competing solvers.
Across these regimes, VeloxQ with default settings usually obtains solution quality and runtime comparable to or better than the tested alternatives, and custom settings allow the runtime-quality trade-off to be adjusted when needed.

For D-Wave quantum annealers, the native Pegasus and Zephyr benchmarks show that VeloxQ obtains solution quality comparable to the QPU on instances that fit directly on the hardware graph.
D-Wave can retain a runtime advantage for some small native instances, especially near the largest native QPU sizes when VeloxQ is run with default settings.
At the same time, the custom VeloxQ settings demonstrate that this gap can be reduced or reversed while maintaining similar solution quality.
The difference becomes more pronounced when the problem topology is not native to the annealer.
For all-to-all instances, D-Wave must rely on embedding, which introduces auxiliary qubits and is associated here with larger reference gaps and a smaller effective problem size.
VeloxQ avoids this embedding step and therefore maintains low reference gaps and shorter runtimes.
For larger sparse Pegasus and Zephyr instances, VeloxQ also extends beyond the range handled by the D-Wave~QPU and Kerberos hybrid solver in this study.

The HUBO benchmarks test whether the auxiliary-variable overhead of reducing higher-order problems to QUBO prevents VeloxQ from remaining useful.
On random three-body Ising chains and weighted MAX-3-SAT instances, VeloxQ remains effective after HUBO-to-QUBO reduction and reaches problem sizes substantially larger than the current gate-based digital-quantum demonstrations considered here.
Within the benchmarked regimes, it also outperforms the available CPLEX and SA references where those references remain practical.
This result does not remove the conceptual advantage of native HUBO solvers, but it shows that, for the tested instances, the scalability of VeloxQ can compensate for the reduction overhead.

The comparisons with CPLEX, brute force, BEIT, and tropical tensor networks probe a different question: how VeloxQ behaves against methods that either provide high-quality references or, in some regimes, certify optimality.
For the Pegasus and Zephyr comparisons with CPLEX, VeloxQ and CPLEX produce comparable solution quality on the smaller instances, but CPLEX reaches the imposed runtime limit as the graph size grows.
Against brute force, BEIT, and TTN, VeloxQ often reaches the same certified energies faster on the tested modest-size instances.
This should be interpreted as a comparison of certification cost against heuristic runtime: VeloxQ can match certified energies in these cases, but it does not provide an optimality guarantee.

The planted-solution benchmarks compare VeloxQ with parallel annealing and simulated bifurcation on instances where the ground-state energy is known by construction.
For 3R3X, tile-planting, and Wishart-planted families, VeloxQ is competitive across both sparse and dense planted instances.
PA can be faster on some small instances, and SBM is strongest on selected easy-regime Wishart instances.
However, VeloxQ remains competitive across the full set of planted families, and custom settings improve the runtime-quality trade-off in several regimes without changing the benchmark instances.
These results are useful because they test solver behavior on controlled hard/easy transitions rather than only on random unstructured inputs.

The B\&B benchmarks compare VeloxQ with search methods designed around explicit bounds and pruning.
On the dense pseudo-random instances considered here, VeloxQ obtains comparable or better reference gaps than the tested B\&B variants and is at least two orders of magnitude faster.
The result is consistent with the broader pattern of the paper: methods with stronger guarantees or more explicit structure can be attractive at modest sizes, but their computational cost grows quickly in regimes where VeloxQ remains practical.

Taken together, the results show that VeloxQ is a scalable QUBO solver with competitive solution quality and runtime across the benchmark families considered in this study.
Its strongest distinction is the ability to operate in ultra-large sparse regimes that the compared quantum and classical methods did not practically reach in this study, while still performing competitively on smaller structured benchmarks.
The main limitation is that VeloxQ is heuristic and does not provide certificates of optimality on its own.
With that qualification, the benchmarks support VeloxQ as a practical solver for large-scale QUBO and QUBO-reduced HUBO problems, particularly when topology flexibility and scalability are central requirements.

\section*{Acknowledgments}
Quantumz.io Sp. z o.o. acknowledges support received from the National Centre for Research and Development (NCBR), Poland, under Project
No. POIR.01.01.01-00-0061/22, titled \textit{VeloxQ: Utilizing Dynamic Systems in Decision-Making Processes Based on Knowledge
    Acquired Through Machine Learning, Across Various Levels of Complexity, in Industrial Process Optimization}, that aims to develop
digital solutions for solving combinatorial optimization problems. The focus is on leveraging quantum-inspired algorithms and artificial
intelligence to enhance decision-making processes in industrial optimization contexts.

\appendix

\section{Transformations between QUBO and Ising model formulations}
\label{app:QUBOIsingconversion}
This appendix follows the conventions of~\cite{Boettcher}. To represent a QUBO problem as an instance of the Ising model, one transforms the binary variables
$x_{i}=\frac{1}{2}\left(1+\sigma_{i}\right)$. In this case the parameters of the related Ising model are given by
\begin{align}
    h_{i}  & =\sum_{j=1}^{N}Q_{ij},\label{eq:Jh_QUBO} \\
    J_{ij} & =\begin{cases}
                  Q_{ij}, & i\not=j, \\
                  \\
                  0,      & i=j,
              \end{cases}
\end{align}
and values of the relevant quadratic functions are related as
\begin{equation}
    Q(\vb{x})=\frac{1}{2}H(\vb{s})-\frac{1}{2}C\label{eq:qubo_ising},
\end{equation}
with $C=\sum_{i=1}^{N}\sum_{j=i+1}^{N}Q_{ij}+\sum_{i=1}^{N}Q_{ii}$.

On the other hand, an Ising model can be represented as a QUBO problem via the discrete variable transformation $s_{i}=2x_{i}-1$.
The parameters of the related QUBO model are given by
\begin{eqnarray}
    Q_{ij} & = & \begin{cases}
        J_{ij},                     & i\not=j, \\
        \\
        h_{i}-\sum_{l=1}^{N}J_{il}, & i=j,
    \end{cases}\label{eq:qij_SK}
\end{eqnarray}
and the relation between the functions is
\begin{equation}
    H(\vb{s})=2Q(\vb{x})+C,\label{eq:H_QUBO}
\end{equation}
with $C=-\sum_{i=1}^{N}\sum_{j=i+1}^{N}J_{ij}+\sum_{i=1}^{N}h_{i}$.

\section{Details about Parallel Annealing and Simulated Bifurcation}
\label{app:solvers}
Simulated annealing (SA) is a well-known heuristic optimization algorithm that employs thermal fluctuations to escape local minima. During the evolution
of the system, the temperature is gradually decreased, leading to decreasing fluctuations and a higher probability of accepting only downhill moves. Ultimately,
the system settles in some low-energy state. On the other hand, Parallel Annealing (PA) is inspired by quantum adiabatic annealing, in which the system's Hamiltonian
is time-dependent, and gradually switches from an initial Hamiltonian with a known ground state to the target Hamiltonian. According to the Adiabatic Theorem~\cite{Born1928} from
quantum mechanics, if this evolution is performed slowly enough, the system will remain in the ground state of the Hamiltonian at all times, thus
reaching the ground state of the target Hamiltonian and solving the optimization problem. PA adapts this idea to the realm of classical heuristic optimization,
by considering a time-dependent, classical Hamiltonian of the form
\begin{equation}
    H(t) = \lambda(t) H_{\text{initial}} + H_{\text{target}},
\end{equation}
where \(t\) plays the role of the time parameter, and \(\lambda(t)\) is a~function that gradually decreases from a sufficiently large value to zero. By analogy
with the quantum case, \(H_{\text{initial}}\) is chosen to be a function with an easily obtainable global minimum. In the original work~\cite{Jiang2023}, the authors
employed a quadratic function \(H_{\text{initial}} = 1/2\sum_{i} x_i^2\). Since there is no actual physical process that would evolve towards low-energy configurations,
a manual update scheme has to be introduced. Intermediate analog spin variables \(x_i\), taking values in the interval \([-1, 1]\), are used as a proxy for the true binary variables.
The classical spins can in turn be retrieved via the sign function, \(s_i = \text{sign}(x_i)\). The update rule for the analog variables is given by a gradient
descent step,
\begin{equation}
    x_i(t+\Delta t) = x_i(t) - \eta \nabla H(t),
\end{equation}
with the gradient computed as a function of continuous variables, but the target part evaluated with binarized spins:
\begin{align}
    \nabla H(t) = & \lambda(t)\nabla H_{\text{initial}} + \nabla H_{\text{target}}  \nonumber \\
    =             & \lambda(t) \vb{x} +  J \vb{s} + \vb{h}.
\end{align}
Drawing inspiration from neural-network training, clipping~\cite{Courbariaux2016},
\begin{equation}
    \vb{x} \to \max(-1, \min(1, \vb{x})),
\end{equation}
and momentum~\cite{Qian1999}
\begin{align}
    \vb{m}(t+1) & =  \alpha \vb{m}(t) - \eta \nabla H(t), \\
    \vb{x}(t+1) & = \vb{x}(t) + \vb{m}(t+1),
\end{align}
are also utilized to stabilize the optimization process.

Another optimization algorithm inspired by quantum adiabatic computing is the Simulated Bifurcation Machine (SBM)~\cite{Goto2019}. It is a classical approximation of
a~network of nonlinear quantum oscillators described by a~chaotic system of Hamilton's equations:
\begin{widetext}
    \begin{align}
        H         & = \frac{a_0}{2} \sum_{i=1}^{N} p_i^2 + \sum_{i=1}^{N} \left(\frac{q_i^4}{4} + \frac{a_0 - a(t)}{2} q_i^2 \right) - c_0
        \sum_{i=1}^{N} \left(h_i q_i + \frac{1}{2}\sum_{j=1}^N J_{ij} q_i q_j\right),                                                                            \\
        \dot{q}_i & = \frac{\partial H}{\partial p_i} = a_0 p_i \label{eq:sbm2},                                                                                 \\
        \dot{p}_i & = -\frac{\partial H}{\partial q_i} = -\left[q_i^2 + a_0 - a(t)\right] q_i + c_0 \left(\sum_{j=1}^{N} J_{ij} q_j + h_i\right). \label{eq:sbm}
    \end{align}
\end{widetext}
This process can be viewed as the dynamics of particles with mass \(a_0^{-1}\), positions \(q_i \in \mathbb{R}\), and momenta \(p_i \in \mathbb{R}\),
in a time-dependent single-particle potential and interacting through Ising-like interactions.
The matrix \(J\) and vector \(h\) represent the Ising minimization problem being solved.
Hyperparameters \(a_0\) and \(c_0\) are typically taken to be \(a_0 = 1\) and \(c_0 = 1/\lambda_{\text{max}}\), where \(\lambda_{\text{max}}\) is the largest
eigenvalue of \(J\) (to bound the extremal values of the Ising term), while \(a(t)\) is a linearly increasing time-dependent function that drives the system through the bifurcation point,
which occurs roughly when \(a(t) = a_0\). After the bifurcation, the system's energy landscape approximately encodes the local minima
of the Ising term, leading the particles to converge toward low-energy solutions of the binary optimization problem.
These solutions can be extracted by taking \(s_i = \mathrm{sign}(q_i)\). Additionally, the chaotic dynamics of the SBM equations
result in sensitivity to initial conditions, allowing many independent replicas to
be run simultaneously from different starting points, further enhancing efficiency and enabling massive parallelization.

\section{Bound functions used in B\&B methods}
\label{app:BB}

\begin{figure}[!tbp]
    \centering
    \includegraphics[width=\linewidth]{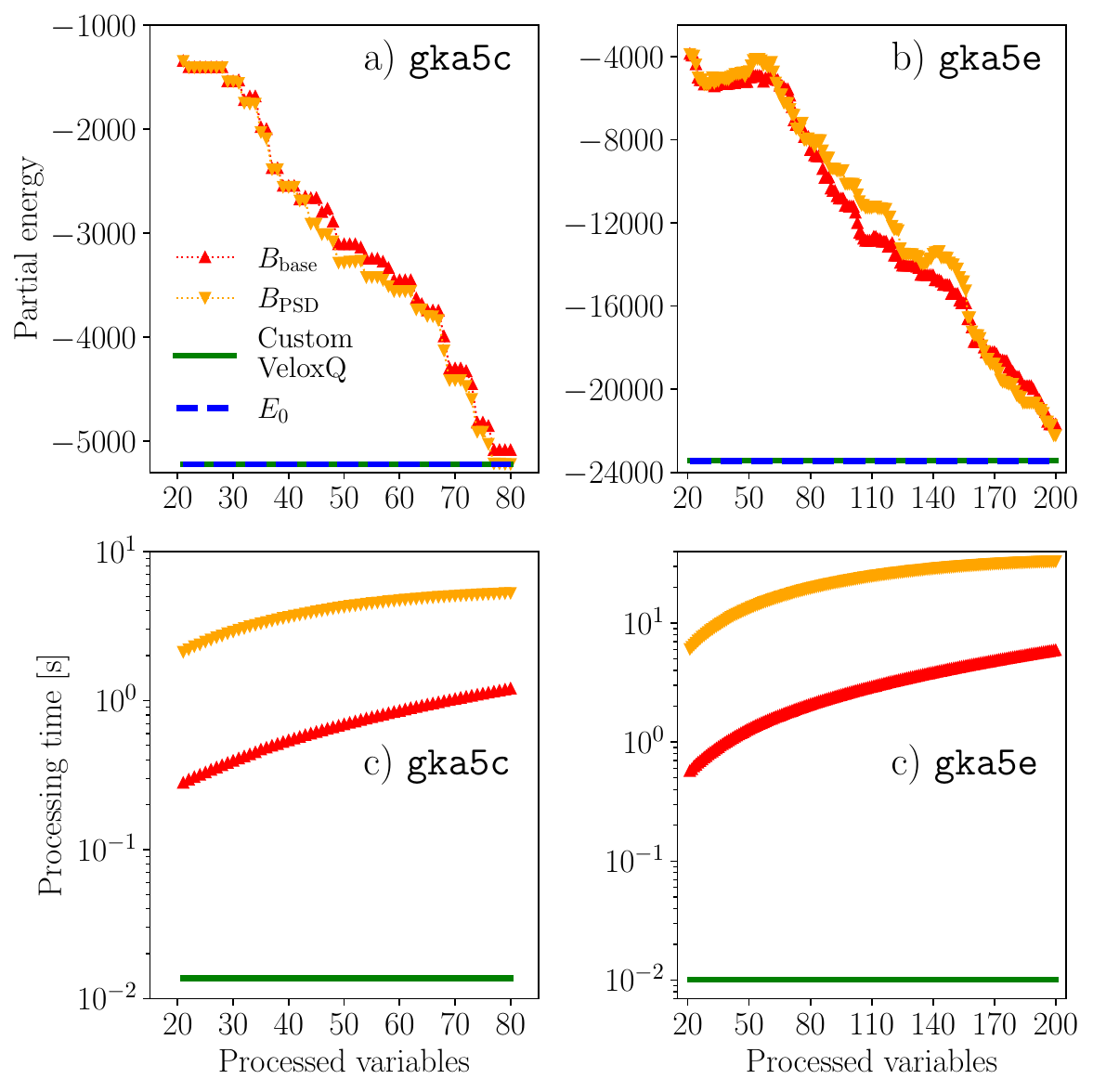}
    \caption{
        The B\&B progress for selected instances demonstrates that the dynamics of $B_{\text{base}}$ and $B_{\text{PSD}}$ are roughly similar.
        $B_{\text{base}}$ achieves better intermediate states for \texttt{gk5e}, but fails to recognize the more optimal state found by $B_{\text{PSD}}$.
        While the time characteristics of $B_{\text{PSD}}$ are slower initially, as the size of the remaining subproblems decreases, the subsequent iterations of
        $B_{\text{PSD}}$ become progressively faster. However, the overall execution time of either B\&B method exceeds the completion time of~VeloxQ early in the run,
        underscoring the performance advantage of VeloxQ on these instances.
    }
    \label{fig:BBGKA}
\end{figure}

The simple B\&B method we adopt utilizes the subproblem energy as the primal bound function. For an Ising model instance defined by $(J, h)$, given a state $s$ and a subproblem $U \subseteq \{1, \ldots, n\}$, the bound is computed as:
\begin{equation}
    B_{\text{base}}(J, h, s, U) = \sum_{i \in U} h_i s_i + \sum_{\substack{i, j \in U \\ i < j}} J_{ij} s_i s_j.
\end{equation}

While this method is efficient when applied to batches of subproblems incrementally, it does not account for the remaining subsets of spins outside $U$. This limitation can be addressed by employing alternative bound functions.

\textbf{PSD problem formulation.} The improved B\&B method involves transforming each evaluated subproblem into a~form based on a positive semidefinite (PSD) matrix. The~relaxed solutions to such subproblems can be used as a bound function in the B\&B algorithm, following the modern algorithmic outline published in~\cite{Quantagonia2023, Chalkis2023}. We refer to a~generalized, matrix-based formulation of the Ising model for $J$ defined as a symmetric connection matrix:
\begin{equation}
    E(J, h, s) = \frac{1}{2} s^T J s + s^T h.
\end{equation}

This formulation can be extended to matrices that describe numerical problems beyond the Ising model, notably with a non-zero diagonal of $J$. In order to reformulate the problem with a PSD matrix, we shift all the diagonal elements by
\begin{equation}
    d(J) = \operatorname{max}(0, -\operatorname{eigmin} J) + \varepsilon,
\end{equation}
where $\varepsilon > 0$ is included to ensure numerical stability. This ensures that $\widetilde{J} = J + d(J) \mathbb{1}$ is a positive definite matrix.

Using $\widetilde{J}$ instead of $J$ introduces an offset to the energy that is constant with respect to $s$
\begin{equation} \label{eq:jjt}
    E(\widetilde{J}, h, s) = E(J, h, s) + d(J) \cdot n,
\end{equation}
which means that the binary problems have the same extrema.

Relaxation of this problem into a continuous domain $r \in \mathbb{R}^n$ leads to the following quadratic PSD form:
\begin{equation}
    \begin{split}
         & E(\widetilde{J}, h, r) = \frac{1}{2} r^T \widetilde{J} r + r^T h ~\rightarrow~ \min.
    \end{split}
\end{equation}
Notably, the equality \eqref{eq:jjt} is ensured only for binary states $r$. While the PSD formulation does not directly solve the original Ising model, it provides a tractable heuristic, resulting in efficient approximation of solutions.

\textbf{Solution.} The relaxed PSD problem can be solved using methods such as proximal operators. The optimal relaxed solution is given by:
\begin{equation}
    r^\ast(J, h) = -\widetilde{J}^{-1} h.
\end{equation}
Since $\widetilde{J}$ is positive definite, it is non-singular, ensuring the existence and uniqueness of $r^\ast(J, h)$.

\vspace{\baselineskip}

\textbf{Bound function for B\&B.} For the complete Ising problem $(J, h)$, the heuristic score can be computed as:
\begin{equation}
    B_{\text{PSD}}(J, h, s) = H(\widetilde{J},~ h,~ -\widetilde{J}^{-1} h).
\end{equation}

In the context of B\&B, the bound function is supposed to estimate the optimal energy of a remaining subproblem. For a subset of spins $U \subseteq \{1, \ldots, n\}$, let $\widetilde{U} = \{1, \ldots, n\} \setminus U$. Then:
\begin{equation}
    B_{\text{PSD}}(J, h, s, U) = B_{\text{PSD}}(J_{\widetilde{U} \times \widetilde{U}}, h_{\widetilde{U}}, s),
\end{equation}
where $J_{\widetilde{U} \times \widetilde{U}}$ and $h_{\widetilde{U}}$ represent the reduced matrices and vectors for the remaining subtrees in the search process.
This approach involves solving a linear system, which can be implemented in batches of states through matrix factorization techniques, such as Cholesky decomposition, chosen based on the sparsity of the $\widetilde{J}$ matrix. Additionally, determining the shift parameter \(d\) requires computing the minimum eigenvalue of the matrix. In our simulations, we employ ARPACK-inspired algorithms~\cite{Lehoucq1998}, adapted for the CUBLAS framework, to ensure efficient computation. Alternative methods tailored to specific runtime environments can make this approach feasible for quantum computing applications~\cite{Brand2022}.

To validate this approach, we also performed benchmarks using the GKA dataset~\cite{Glover1998}, a recognized standard for evaluating B\&B methods for quadratic problems~\cite{Kochenberger2014}, included in the Biq Mac Library~\cite{BiqMacLibrary}. Files and optimal energies from the Biq Mac Library were converted to Ising formulations, and the benchmarked B\&B methods were configured to maintain $2^{20}$ states in memory. Using the $B_{\text{PSD}}$ bound function resulted in execution times that were 2 to 6 times longer than the baseline B\&B method, but it achieved smaller energy gaps in 75\% of cases, especially for larger instances. VeloxQ used a simple setup with 150 internal states and AutoTune, achieving lower energy gaps with significantly lower runtime on most instances. One instance from the set (\texttt{gka1d}) was solved using B\&B with $B_{\text{PSD}}$, while VeloxQ did not reach the ground state in that case. The results of these benchmarks are presented in Table~\ref{tbl:BBGKA} and Fig.~\ref{fig:BBGKA}.

\begin{table}[!tbp]
    \centering
    \begin{tabular}{|c|c||c|c||c|c||c|c|}
        \cline{2-8} \multicolumn{1}{c|}{} & \multirow{2}{*}{$\mathbf{N}$} & \multicolumn{2}{|c|}{$\mathbf{B_{\text{base}}}$} & \multicolumn{2}{|c|}{$\mathbf{B_{\text{PSD}}}$} & \multicolumn{2}{|c|}{\textbf{VeloxQ}}                                                                       \\
        \cline{3-8} \multicolumn{1}{c|}{} &                               & \textbf{gap}                                     & $\mathbf{t~[\mbox{s}]}$                         & \textbf{gap}                          & $\mathbf{t~[\mbox{s}]}$ & \textbf{gap}    & $\mathbf{t~[\mbox{s}]}$ \\ \hline\hline
        \texttt{gka1a}                    & 50                            & \textbf{0.0\%}                                   & 0.41                                            & \textbf{0.0\%}                        & 0.98                    & \textbf{0.0\%}  & 0.01                    \\ \hline
        \texttt{gka2a}                    & 60                            & \textbf{0.0\%}                                   & 0.68                                            & \textbf{0.0\%}                        & 1.71                    & \textbf{0.0\%}  & 0.01                    \\ \hline
        \texttt{gka3a}                    & 70                            & 0.7\%                                            & 1.18                                            & 0.7\%                                 & 2.6                     & \textbf{0.0\%}  & 0.01                    \\ \hline
        \texttt{gka4a}                    & 80                            & \textbf{0.0\%}                                   & 1.1                                             & \textbf{0.0\%}                        & 3.44                    & \textbf{0.0\%}  & 0.01                    \\ \hline
        \texttt{gka5a}                    & 50                            & \textbf{0.0\%}                                   & 0.44                                            & \textbf{0.0\%}                        & 1.0                     & \textbf{0.0\%}  & 0.01                    \\ \hline\hline
        \texttt{gka1c}                    & 40                            & \textbf{0.0\%}                                   & 0.28                                            & \textbf{0.0\%}                        & 0.5                     & \textbf{0.0\%}  & 0.01                    \\ \hline
        \texttt{gka2c}                    & 50                            & \textbf{0.0\%}                                   & 0.5                                             & \textbf{0.0\%}                        & 1.1                     & \textbf{0.0\%}  & 0.01                    \\ \hline
        \texttt{gka3c}                    & 60                            & 1.85\%                                           & 0.66                                            & \textbf{0.0\%}                        & 1.57                    & \textbf{0.0\%}  & 0.01                    \\ \hline
        \texttt{gka4c}                    & 70                            & \textbf{0.0\%}                                   & 0.88                                            & 0.92\%                                & 2.43                    & \textbf{0.0\%}  & 0.01                    \\ \hline
        \texttt{gka5c}                    & 80                            & 2.62\%                                           & 1.04                                            & 0.08\%                                & 3.31                    & \textbf{0.0\%}  & 0.01                    \\ \hline\hline
        \texttt{gka1d}                    & 100                           & 4.34\%                                           & 1.53                                            & \textbf{0.0\%}                        & 5.72                    & 0.24\%          & 0.01                    \\ \hline
        \texttt{gka2d}                    & 100                           & 2.27\%                                           & 1.6                                             & 7.73\%                                & 6.01                    & \textbf{0.0\%}  & 0.01                    \\ \hline
        \texttt{gka3d}                    & 100                           & 1.77\%                                           & 1.74                                            & 1.28\%                                & 5.9                     & \textbf{0.33\%} & 0.01                    \\ \hline
        \texttt{gka4d}                    & 100                           & 5.48\%                                           & 1.65                                            & 2.5\%                                 & 5.82                    & \textbf{0.0\%}  & 0.01                    \\ \hline
        \texttt{gka5d}                    & 100                           & 3.16\%                                           & 1.57                                            & 0.64\%                                & 5.8                     & \textbf{0.0\%}  & 0.01                    \\ \hline\hline
        \texttt{gka1e}                    & 200                           & 1.17\%                                           & 5.41                                            & 10.9\%                                & 27.6                    & \textbf{0.34\%} & 0.05                    \\ \hline
        \texttt{gka2e}                    & 200                           & 11.17\%                                          & 15.41                                           & 6.76\%                                & 47.51                   & \textbf{0.0\%}  & 0.01                    \\ \hline
        \texttt{gka4e}                    & 200                           & 5.72\%                                           & 5.57                                            & 3.8\%                                 & 27.8                    & \textbf{0.0\%}  & 0.01                    \\ \hline
        \texttt{gka5e}                    & 200                           & 7.5\%                                            & 5.49                                            & 5.28\%                                & 27.53                   & \textbf{0.03\%} & 0.01                    \\ \hline
    \end{tabular}
    \caption{Computational results for selected \texttt{GKA} instances~\cite{Glover1998}.}
    \label{tbl:BBGKA}
\end{table}

\bibliography{lit}

\end{document}